\documentclass{article}
\usepackage{epsfig}

\date{\today}

\usepackage{amsmath}

\newcommand{\ee}{\end{equation}}
\newcommand{\eea}{\end{eqnarray}}
\newcommand{\be}{\begin{equation}}
\newcommand{\bea}{\begin{eqnarray}}

\begin{document}

\title{Black strings 
in $(4+1)-$dimensional \\ Einstein-Yang-Mills theory}
\author{{\large Yves Brihaye \footnote{yves.brihaye@umh.ac.be}}\\
\small{
Facult\'e des Sciences, Universit\'e de Mons-Hainaut,
B-7000 Mons, Belgium }\\
{ }\\
{\large Betti Hartmann \footnote{b.hartmann@iu-bremen.de}}\\
\small{
School of Engineering and Science, International University Bremen,
28725 Bremen, Germany
 }\\
{ }\\
{\large Eugen Radu\footnote{radu@thphys.nuim.ie}}\\
\small{
Department of  Mathematical Physics,
National University of Ireland Maynooth, Ireland}}

\date{\today}

\maketitle
\begin{abstract}
We study two classes of static uniform black string solutions
in a $(4+1)$-dimensional SU(2) Einstein-Yang-Mills model. 
These configurations possess a regular event horizon and 
corresponds in a $4-$dimensional
picture to axially symmetric black hole
solutions in an Einstein-Yang-Mills-Higgs-U(1)-dilaton theory.
In this approach, one set of solutions possesses a nonzero magnetic charge,
while the other solutions represent black holes
located in between a monopole-antimonopole pair.
A detailed analysis of the solutions' properties
is presented, the domain of existence of the black strings being determined.
New four dimensional solutions are found 
by boosting the five dimensional configurations.
We also present an argument for the non-existence 
of finite mass hyperspherically symmetric 
black holes in SU(2) Einstein-Yang-Mills theory.  
 
\end{abstract}

\section{Introduction}
Black holes in more than four spacetime dimensions have a much richer
spectrum of horizon topologies than their four dimensional counterparts.
In a $d-$dimensional asymptotically flat spacetime, the static vacuum 
black hole of a certain mass with 
horizon topology  $S^{d-2}$ is
uniquelly described  by the Schwarzschild-Tangherlini solution \cite{tang}.
While this solution is hyperspherically symmetric, so-called black strings
also exist \cite{bs}. 
Such configurations are important if one supposes 
the existence of extra dimensions in the universe, 
which are likely to be compact and described by a Kaluza-Klein (KK) theory.
The simplest vacuum static solution of this 
type (and the only one known in closed form)
is found by assuming translational symmetry along the 
extra coordinate direction and corresponds to a uniform black string
with horizon topology $S^{d-3}\times S$. 
Though this solution exists for all values of the mass, 
it is unstable below a critical value
as shown by Gregory and Laflamme \cite{Gregory:1993vy}.
This was interpreted to mean that a light uniform string decays
to a (hyperspherical) black hole since that has higher entropy.
However, Horowitz and Maeda \cite{Horowitz:2001cz} argued
that such a transition has an intermediate step: the light uniform
string decays to a non-uniform string, which then eventually decays
to a black hole.
This prompted a search for this missing link, and a branch of
non-uniform black string solutions was found in
\cite{Gubser:2001ac,Wiseman:2002zc}.
Nevertheless, a number of aspects still remains to be clarified
and 
the literature on non-uniform black string solutions  is continuously
growing 
(see \cite{Harmark:2005pp} for a recent review).

The uniform black string solutions have been 
explored from many points of view, in particular
by including various matter fields.
However, most of the studies in the literature restricted to 
the case of an abelian matter content.
At the same time, a number of results obtained in the literature clearly
indicates that the solutions of Einstein's equations coupled to nonabelian
matter fields possess a much richer structure 
than in the U(1) case (see \cite{Volkov:1998cc} for a survey of the situation in $d=4$).
Here the only case systematically discussed in the literature is $d=5$ and an SU(2) gauge field, 
for an ansatz with no dependence on the extra $z-$coordinate.
No exact solutions with reasonable asymptotics 
are available analytically in this case and the field equations 
have to be solved numerically.
As a new feature, both globally regular 
and black string solutions are possible to exist.
The simplest vortex-type configurations have been constructed 
in \cite{Volkov:2001tb} and
are spherically symmetric in four dimensions, extending
trivially into one extra dimension. 
The  black string counterparts of these solutions have been constructed 
recently in \cite{Hartmann:2004tx} and found to present a complicated branch structure.

However, as argued in \cite{Brihaye:2005pz},
the existence of a nonabelian winding number implies a very rich
set of possible boundary conditions, the configurations which are spherically
symmetric in four dimensions corresponding to the simplest case.
A discussion of a more general set of 
vortex-type, globally regular solutions have been presented
in \cite{Brihaye:2005pz}. In a  four dimensional picture these correspond to axially
symmetric multimonopoles (MM), respectively monopole-antimonopole (MA)
solutions in an Einstein-Yang-Mills-Higgs (EYMH) theory with a 
non-trivial coupling to  
a Maxwell and a dilaton field.
A preliminary discussion of the black string counterparts 
of some of these solutions
has been presented in \cite{Brihaye:2005fz}.

It is the purpose of this paper to present a systematic analysis
of the uniform black string solutions in Einstein-Yang-Mills (EYM) 
theory in five spacetime dimensions, 
for two different sets
of boundary conditions satisfied by the matter fields at infinity.
Apart from the solutions discussed in \cite{Hartmann:2004tx},
we present a detailed study of the 
 deformed  black string
configuration, which in the (3+1)-dimensional reduced theory
can be interpreted as describing axially symmetric multimonopole 
black holes, respectively
nonabelian black holes with magnetic dipole hair.
Although all fields are independent on the extra coordinate, 
the event horizon of these solutions is deformed
(the horizon circumference along the equator differs
from the corresponding quantity evaluated along the poles),
which resembles the non-uniform black string case \cite{Gubser:2001ac,Wiseman:2002zc}.

Although the no hair conjecture is known to be violated for 
$d=5$ solutions with one compact extra dimension 
(see $e.g.$ the discussion in \cite{Horowitz:2002dc})
the solutions presented in this paper provide
another counterexample to this conjecture.
However, this is not a surprise, given the relation noticed above with 
the (3+1)-dimensional theory, which is known to
present hairy black hole solutions.

As already noticed in \cite{Brihaye:2005fz},
after boosting the $d=5$ configurations and reducing to $d=4$,
 we obtain new four dimensional solutions which rotate
 and present a nonzero nonabelian electric charge.
For the particular case of configurations
which are spherically symmetric in four dimensions,
this procedure generates $d=4$ dyonic black holes. 

We argue here that these solutions are interesting from yet another point of view.
Different from other known theories,
no $d=5$ hyperspherically symmetric  finite mass EYM solutions
exist in five dimensions,  unless one considers
the inclusion of higher order curvature 
terms in the action \cite{Brihaye:2002jg}.
The nonexistence proof for the globally regular case
has been presented in \cite{Volkov:2001tb} (see also \cite{Okuyama:2002mh}). 
In Appendix A, we 
present a similar argument for black hole solutions.
Therefore, the simplest $d=5$ solutions of the EYM theory
corresponds to uniform vortices and black strings.

Our paper is organized as follows: In Section 2, we give the model
including the Ansatz and the boundary conditions. 
In Section 3, we describe our numerical results. 
The properties of the  vortex solutions are also reviewed, a number
of new results being presented.
In Section 4, we comment on the new $d=4$ solutions that we
obtain by boosting the $d=5$ EYM configurations and in Section 5 we give our
conclusions.
 
Most of the notations and sign conventions used in this paper are similar to
those in Ref.~\cite{Brihaye:2005pz}.

\section{The model}
\subsection{Action principle}
We consider the five dimensional SU(2) Einstein-Yang-Mills (EYM)  action
\begin{equation}
\label{action5}
I_5=\int d^{5}x\sqrt{-g_m }\Big(\frac{R }{16\pi G}
-\frac{1}{2g^2}Tr\{F_{MN }F^{MN} \}\Big),
\end{equation}
(throughout this paper, the indices $\{M,N,...\}$ 
will denote the five dimensional
coordinates and $\{\mu,\nu,...\}$ the
coordinates of the four dimensional
physical spacetime).

Here $G$ is the gravitational constant,
$R$ is the Ricci scalar associated with the
spacetime metric $g_{MN}$
and
$F_{MN}=\frac{1}{2} \tau^aF_{MN}^{(a)}$ is the gauge field strength
tensor defined as
$F_{MN} =
\partial_M A_N -\partial_N A_M + i[A_M , A_M  ],
$
where the gauge field is
$A_{M} = \frac{1}{2} \tau^a A_M^{(a)},$
$\tau^a$ being the Pauli matrices and $g$ the gauge coupling
constant.

Variation of the action (\ref{action5})
 with respect to  $g^{MN}$ and $A_M$ leads to the field equations
\begin{eqnarray}
\label{einstein-eqs}
R_{MN}-\frac{1}{2}g_{MN}R   &=& 8\pi G  T_{MN},
\\
\label{YM-eqs}
\nabla_{M}F^{MN}+i[A_{M},F^{MN}]&=&0,
\end{eqnarray}
where the YM stress-energy tensor is
\begin{eqnarray}
\label{tik}
T_{MN} = 2{\rm Tr}
    ( F_{MP} F_{NQ} g^{PQ}
   -\frac{1}{4} g_{MN} F_{PQ} F^{PQ}).
\end{eqnarray}

\subsection{The ansatz}
In what follows we will consider uniform black string configurations,
assuming that both the matter functions and
the metric functions are
independent on the extra coordinate $x^5 \equiv z$.
 Without any loss of generality, we consider a five-dimensional
metric parametrization
\begin{eqnarray}
\label{metrica}
ds^2 = e^{- a\psi }\gamma_{\mu \nu}dx^{\mu}dx^{\nu}
 + e^{ 2a\psi }(dz + 2{\cal W}_{\mu}dx^{\mu})^2,
\end{eqnarray}
with $a=2/\sqrt{3}$.

The four dimensional reduction of this theory with respect to the Killing vector
$\partial/\partial z$ has been presented in \cite{Brihaye:2005pz}.
For the  reduction of the YM action term,
a convenient SU(2) ansatz is
\begin{eqnarray}
\label{SU2}
A={\cal A}_{\mu}dx^{\mu}+g\Phi (dz+2 {\cal W}_\mu dx^\mu),
\end{eqnarray}
where ${\cal W}_\mu$ is a U(1) potential,
${\cal A}_{\mu}$ is a purely four-dimensional gauge field potential,
while  $\Phi$ corresponds after the dimensional reduction to a
triplet Higgs field.

This leads to the four dimensional action principle
\begin{eqnarray}
\label{action4}
I_4=\int d^{4}x\sqrt{-\gamma }\Big[
\frac{1}{4\pi G}\big(
\frac{\mathcal{R} }{4}
-\frac{1}{2}\nabla_{\mu}\psi \nabla^{\mu}\psi
-e^{2\sqrt{3}\psi}\frac{1}{4}G_{\mu \nu }G^{\mu \nu } \big)
-e^{2\psi/\sqrt{3}}\frac{1}{2g^2}Tr\{
{\cal F}_{\mu \nu }{\cal F}^{\mu \nu }\}
\\
\nonumber
-e^{-4\psi/\sqrt{3}}Tr\{ D_{\mu}\Phi D^{\mu}\Phi\}
- 2 e^{2\psi/\sqrt{3}}\frac{1}{g}G_{\mu \nu} Tr\{\Phi {\cal F}^{\mu
\nu} \}
-2e^{2\psi/\sqrt{3}} G_{\mu\nu}G^{\mu\nu}Tr\{  \Phi^2 \}
\Big],
\end{eqnarray}
where $\mathcal{R}$ is the Ricci scalar for the metric $\gamma_{\mu
\nu}$,
while
 ${\cal F}_{\mu \nu }=
\partial_{\mu}{\cal A}_{\nu}
-\partial_{\nu}{\cal A}_{\mu}+i [{\cal A}_{\mu},{\cal A}_{\nu}  ]$
and
 $G_{\mu \nu}=\partial_{\mu}{\cal W}_{\nu}-\partial_{\nu}{\cal
W}_{\mu}$
are the SU(2) and U(1) field strength tensors defined in $d=4$.

Here we consider five dimensional configurations possessing two more
Killing vectors
apart from $\partial/\partial z$. These are
$\xi_1=\partial/\partial \varphi$,
corresponding to an axial
symmetry of the four dimensional metric sector
(where the azimuth angle $\varphi$
 ranges from $0$ to $2 \pi$),
and $\xi_2=\partial/\partial t$,
with $t$ the time coordinate.

We consider the following parametrization of the four dimensional
line element, employed also to find globally regular solutions
\begin{equation}
\label{metric}
d\sigma^2=\gamma_{\mu \nu}dx^{\mu}dx^{\nu}=\gamma_{tt}dt^2+d\ell^2=
- f(r,\theta)dt^2 +  \frac{q(r,\theta)}{f(r,\theta)}
(d r^2+ r^2 d \theta^2 )
           +  \frac{l(r,\theta)}{f(r,\theta)} r^2 \sin ^2 \theta
d\varphi^2,
\end{equation}
and the function $g_{zz}$ depending also on $r,\theta$ only.
Here $t$ is the time coordinate, $r$ is the radial coordinate 
while $0\leq \theta< \pi$ is a polar angle.

We take the event horizon to reside at a surface of constant radial coordinate
$r=r_h>0$, characterized by the condition $f(r_h)=0$, $i.e.$ $g_{tt}=0$.
The remaining metric potentials take nonzero and finite values at the event horizon.

The construction of the corresponding YM ansatz  compatible with 
the spacetime symmetries
has been discussed in \cite{Brihaye:2005pz}.
For the time and extra-direction
translational symmetry, we choose a gauge such that
$\partial A/\partial t=\partial A/\partial z=0$.
However, the
action of the Killing vector $\xi_1$
can be compensated by a gauge rotation 
${\mathcal{L}}_{\varphi} A_{N}=D_{N}\bar{\psi},$
with $\bar{\psi}$ being a Lie-algebra valued gauge function \cite{Forgacs:1980zs}.
According to the standard analysis, 
this introduces a winding number $n$ in the ansatz (which is a
constant of motion
and is restricted to be an integer). 

As discussed in  \cite{Brihaye:2005pz}, the
 most general five dimensional  Yang-Mills ansatz compatible with these symmetries
contains 15 functions: 12 magnetic
and 3 electric potentials 
\begin{eqnarray}
\label{A-gen-sph}
A_{N}=\frac{1}{2}A_{N}^{(r)}(r,\theta)\tau_{r}^n
        +\frac{1}{2}A_{N}^{(\theta)}(r,\theta)\tau_{\theta}^n
        +\frac{1}{2}A_{N}^{(\varphi)}(r,\theta)\tau_{\varphi}^n,
\end{eqnarray}
where $\tau_r^n$, $\tau_{\theta}^n$ and
$\tau_{\varphi}^n$ denote the scalar product of the vector of Pauli
matrices $\vec{\tau}=(\tau_1,\tau_2,\tau_3)$ with the unit vectors
$\vec{e}_r^n=(\sin\theta\cos n\varphi, \sin\theta \sin n\varphi, \cos \theta)$,
$\vec{e}_{\theta}^n=(\cos\theta\cos n\varphi, \cos\theta \sin n\varphi, -\sin \theta)$,
$\vec{e}_{\varphi}^n=(-\sin n\varphi, \cos n\varphi, 0)$.

Searching for solutions within the most general ansatz
is a difficult task.
Therefore, similar to \cite{Brihaye:2005pz} we use a purely magnetic reduced ansatz with
six essential nonabelian potentials and
\begin{eqnarray}
\label{ansatz-spec}
\nonumber
A_{r}^{(r)}=A_{r}^{(\theta)}~=~A_{\theta}^{(r)}~=~A_{\theta}^{(\theta)}
~=~A_{\varphi}^{(\varphi)}=~A_{5}^{(\varphi)}=~A_{t}^{(a)}=0.
\end{eqnarray}
The consistency of this reduction has been verified at the level of the
YM equations.

However, for configurations with a nontrivial $\theta-$dependence
the gauge potentials
$A_{\varphi},~A_5$ have components along the same directions in
isospace, which implies that the $T_{\varphi 5}$ component of the
energy-momentum tensor will be nonzero  \cite{Brihaye:2005pz}.
Thus the Einstein equations imply the existence, in the five dimensional metric ansatz
(\ref{metrica}),
of one extradiagonal  $g_{5\varphi}$ metric function, $i.e.$
\begin{eqnarray}
{\cal W}_{\mu}=J(r,\theta)\delta_{\mu}^{\varphi}.
\end{eqnarray}

A suitable parametrization of the  nonzero
components of $A_N^{(a)}$ which factorizes the trivial
$\theta$-dependence
and admits a straightforward four dimensional picture is:
\begin{eqnarray}
\label{ansatz}
&&A_r^{(\varphi)}=\frac{1}{r}H_1(r,\theta),~~~A_{\theta}^{(\varphi)}=1-H_2(r,\theta),
~~~A_{\varphi}^{(r)}=-n \sin \theta H_3(r,\theta)+2g
J(r,\theta)\Phi_1(r,\theta),
\\
\nonumber
&&A_{\varphi}^{(\theta)}=-n \sin \theta (1-H_4(r,\theta))+2g
J(r,\theta)\Phi_2(r,\theta),
~~~A_{5}^{(r)}=\Phi_1(r,\theta),~~~A_{5}^{(\theta)}=\Phi_2(r,\theta).
\end{eqnarray}
The gauge invariant quantities expressed in terms of these functions
will be independent on the
angle $\varphi$. 

To fix the residual abelian gauge invariance we choose the
gauge condition
\begin{eqnarray}
\label{gauge}
\nonumber
r \partial_r H_1 - \partial_\theta H_2 = 0.
\end{eqnarray}

The $d=5$ EYM configurations  extremize also the
action principle (\ref{action4}) and can be viewed
as  solutions of the four dimensional
theory.
In this picture, $H_i(r,\theta)$ are the
 magnetic SU(2) gauge potentials, $\psi(r,\theta)$ is a dilaton,
$J(r,\theta)$
 is a U(1) magnetic potential,
 while $\Phi_1(r,\theta),~\Phi_2(r,\theta)$ are the components of a
Higgs field.
We mention also that, similar to the pure (E)-YMH case, one may define
a 't Hooft field strength tensor and an expression for the nonabelian
electric and magnetic charges
within the action principle (\ref{action4}).

\subsection{Particular solutions}
Restricting to configurations which are spherically symmetric in $d=4$,
one exact solution of the $d=5$ EYM equations
is found taking the product of the
$d=4$ Schwarzschild black hole with a circle
(here we consider an isotropic coordinate system)
\begin{eqnarray} 
\label{m1}
ds^2=(1+\frac{r_h}{2r})^4\left(dr^2+r^2d\Omega^2\right)-\left(\frac{1-\frac{r_h}{2r}}{1+\frac{r_h}{2r}}
\right)^2dt^2+dz^2,
\end{eqnarray}
and for a pure gauge SU(2) field
$
H_1=H_3=\Phi_2=0$, $\Phi_1=const.$,~$H_2=H_4=1$, the event horizon being
located at $r=r_h/2 >0$.

The second solution is more important and corresponds to an embedded U(1) configuration
with 

\begin{eqnarray} 
\label{m2}
ds^2=\left(\frac{r^2+2(r_++r_-)r+(r_+-r_-)^2}{4 r^2}\right)^2
\left(dr^2+r^2d\Omega^2\right)-
\left(
1-\frac{4r_+r}{r^2+2(r_++r_-)r+(r_+-r_-)^2}
\right) dt^2
\\
\nonumber
+
\left(
1-\frac{4r_-r}{r^2+2(r_++r_-)r+(r_+-r_-)^2}
\right)dz^2,
\end{eqnarray}
and
$
H_i=\Phi_2=0$, $\Phi_1=const.$, where
$r_+$ and $r_-$ are two constants with $r_+r_-=4\alpha^2 n^2/3$. 
In a four dimensional perspective, this describes a Dirac monopole
in an Einstein-Maxwell-dilaton theory (the $U(1)$ field
originating here from the $d=5$ nonabelian field), with a dilaton coupling constant $1/\sqrt{3}$
(see e.g. \cite{Garfinkle:1990qj}).
These solutions have an
regular event horizon at $r=r_+ - r_-$. There is also a curvature singularity at $r=r_-$.
The physical mass of the solution is 
$M=(2r_++r_-)/4$, while the magnetic charge is $Q_M=\sqrt{3r_+r_-}/2$.
\subsection{Boundary conditions}
\subsubsection{Metric functions}
To obtain finite energy density black string solutions
 that asymptote
to $M^4\times S$, where $M^4$ is the four dimensional Minkowski spacetime,
the metric functions have to satisfy the boundary conditions
\begin{equation} 
\label{b1}
\partial_r \psi|_{r=r_h}=f|_{r=r_h}= q|_{r=r_h}=
l|_{r=r_h}= \partial_r J|_{r=r_h}=0,
\end{equation}
on the event horizon, and
\begin{equation}\label{b2}
f|_{r=\infty}= q|_{r=\infty}=
l|_{r=\infty}=1,~\psi|_{r=\infty}=~J|_{r=\infty}=0,
\end{equation}
at infinity.
For solutions with parity reflection symmetry (the case considered in
this paper),
the boundary conditions along the $\bar{z}$ and $\bar{\rho}$ axes are
(with $\bar{z}=r \cos \theta$ and $\bar{\rho}=r \sin \theta$)
\begin{equation}
\partial_\theta \psi|_{\theta=0,\pi/2}=
\partial_\theta J|_{\theta=0,\pi/2}=
\partial_\theta f|_{\theta=0,\pi/2} =
\partial_\theta q|_{\theta=0,\pi/2} =
\partial_\theta l|_{\theta=0,\pi/2} =0.
\end{equation} 

\subsubsection{Matter functions}

In \cite{Brihaye:2005pz} the following general 
set of boundary conditions at $r\to \infty$
has been proposed for the magnetic potentials $\Phi_1$ and $\Phi_2$
\begin{equation}
\label{phi-rinfty}
\lim_{r\to\infty}\Phi_1=\eta \cos m\theta,~~~
\lim_{r\to\infty}\Phi_2=\eta \sin m\theta\,,
\end{equation}
with $m=0,1,\dots$, and $\eta$ an arbitrary constant.

The expression of the boundary conditions satisfied by the gauge functions $H_i$
at $r\to \infty$
depends on the value of $m$ and
is given in \cite{Brihaye:2005pz}.
In this paper we restrict ourselves to the simplest cases, $m=0$ and
$m=1$, corresponding in a four dimensional picture to multimonopoles (MM)
and monopole-antimonopole (MA) configurations, respectively.
For $m=0$ one finds
\begin{equation}
\label{H-rinfty1}
H_i|_{r=\infty}=0,
\end{equation}
with $i=1,4$, while the $m=1$ solutions are found within a set of boundary conditions
\begin{equation}
\label{H-rinfty2}
H_1|_{r=\infty}=H_2|_{r=\infty}=0,~~H_2|_{r=\infty}=H_4|_{r=\infty}=-1.
\end{equation}
In both cases, the boundary values at the event horizon are 
\begin{eqnarray}
\label{rhMM}
H_{1}|_{r=r_h}=0, ~~
\partial_r H_{2}|_{r=r_h}=\partial_r H_{3}|_{r=r_h}=\partial_r H_{4}|_{r=r_h}=0, ~~
\partial_r \Phi_{1}|_{r=r_h}=\partial_r \Phi_{2}|_{r=r_h}=0,
\end{eqnarray} 
  The conditions along the axes are determined by the symmetries and 
finite energy density requirements. 
For $m=0$ solutions we impose  
\begin{equation}
H_1|_{\theta=0,\pi/2}=H_3|_{\theta=0,\pi/2}=\Phi_2|_{\theta=0,\pi/2}=0,~~
\partial_\theta H_2|_{\theta=0,\pi/2}
= \partial_\theta H_4|_{\theta=0,\pi/2}= \partial_\theta
\Phi_1|_{\theta=0,\pi/2}
= 0,
\end{equation}
which in a four dimensional picture implies a magnetic charge $Q_M=n$.
The conditions satisfied by the $m=1$  configurations are
\begin{eqnarray}
\label{asm1}
H_1|_{\theta=0,\pi/2}=H_3|_{\theta=0,\pi/2}=
\partial_\theta H_2|_{\theta=0,\pi/2}=\partial_\theta
H_4|_{\theta=0,\pi/2}=0,
\\
\nonumber
\partial_\theta \Phi_1|_{\theta=0}=\Phi_1|_{\theta=\pi/2}=
\Phi_2|_{\theta=0}=\partial_\theta \Phi_2|_{\theta=\pi/2}=0.
\end{eqnarray}
In addition, regularity on the $\bar{z}-$axis requires the conditions $l|_{\theta=0}=q|_{\theta=0}$,
 $H_2|_{\theta=0}=H_4|_{\theta=0}$ to be satisfied, for any values of the integers $(m,~n)$.

 \subsection{Physical parameters} 
The physical parameters of the solution can be computed by taking either a
five-dimensional viewpoint, or in terms of the four-dimensional theory that is
obtained via KK reduction. Both kind of viewpoints are of course directly related, 
but the five-dimensional theory is simpler and so we take this viewpoint in this section.

For a given solution we consider the asymptotic region as
defined by $r \rightarrow \infty$.
The field equations imply the following asymptotic
form of the relevant metric potentials
$g_{tt}$ and $g_{zz}$ 
\begin{equation}
\label{asympt1}
g_{tt} \simeq -1 + \frac{c_t}{r }+O(1/r^2),
~~~
g_{zz} \simeq 1 + \frac{c_z}{r }+O(1/r^2) \ ,
\end{equation}
while the functions which enter the four dimensional line-element (\ref{metric})
behave for large $r$ as 
\begin{eqnarray}
\label{a21}
f\simeq 1+\frac{f_1}{r}+O(1/r^3),~~
q\simeq1+\frac{q_1+q_2 \sin^2 \theta}{r^2}+O(1/r^3),~~l\simeq 1+\frac{q_1}{r^2}+O(1/r^3),
\end{eqnarray}
where $c_t,~c_z$, $f_1$, $q_1$ and $q_2$ are arbitrary real constants 
(with $f_1=c_z/2-c_t$).

Here it is convenient to use the general formalism proposed in \cite{Lu:1993vt} 
(see also \cite{Myers:1999ps}).
The $d=5$ spacetime has a translation-invariant direction $z$ and hence
it can be assigned a $2\times2$ ADM stress-energy tensor $T_{ab}$,
$a,b=t,z$. 
This is computed from the asymptotic metric
$g_{MN}=\eta_{MN}+h_{MN}$ (in
Cartesian coordinates) as
\be
T_{ab}=\frac{1}{16\pi G}\int d\Omega_{(2)}r^2
n^i\left[\eta_{ab}\left(\partial_i h^c_c +\partial_i h^j_j -\partial_j h^j_i\right)-
\partial_i h_{ab}\right)]
\ee
where $n^i$ is the radial normal vector and $a,b,c$ run over parallel
directions $t,z$ while $i,j$ run over transverse directions.
The integration is over the transverse angular directions. Using this we obtain
the mass and momentum as the integrated energy and momentum
densities.

In terms of $c_t$, $c_z$ the mass $M$
and the tension $\tilde{\sigma}$ along the $z$-direction are given by
\begin{equation}
M = \frac{ L}{4 G }
\left[ 2 c_t - c_z \right],
~~~
\tilde{\sigma} = \frac{1}{4 G }
\left[ c_t - 2 c_z \right] \ ,
\end{equation}
where $L$ is the length of the extra dimension, which - if not stated differently -
is set to one in this paper.
Note that a similar mass expression is found by applying 
the Hawking-Horowitz mass formula \cite{Hawking:1995fd}.
Another relavant quantity here is the relative tension \cite{Harmark:2003dg}
defined as
\begin{equation}
\label{reltens} 
n_t=\frac{\tilde{\sigma}L}{M}=\frac{c_t-2c_z}{2c_t-c_z},
\end{equation}
with $0 \leq n_t \leq 2$ for any $d=5$ static solution.
A vacuum uniform black string has $n_t=1/2$ and exists for all values of $M$,
while in the nonuniform case, knowing the exact curve of a branch in 
the $(M,n_t)$ phase diagram enables one to obtain the entire thermodynamics
of that branch.

\subsection{Temperature, entropy and deformation of the horizon}
The zeroth law of black hole physics states that 
the surface gravity $\kappa$ 
is constant at the horizon of the black hole solutions,
where 
\begin{equation}
\label{kappa} 
\kappa^2 =
-(1/4)g^{tt}g^{ij}(\partial_i g_{tt})(\partial_j g_{tt})\Big|_{r=r_h}. 
\end{equation}
To evaluate $\kappa$, we 
use the following expansions of the metric functions at the horizon
\begin{eqnarray}
\label{expan-h}
\nonumber
f(r,\theta)&=&f_2(\theta)\left(\frac{r-r_h}{r_h}\right)^2 
 + O\left(\frac{r-r_h}{r_h}\right)^3 \ ,
\\
q(r,\theta)&=&q_2(\theta)\left(\frac{r-r_h}{r_h}\right)^2 
 + O\left(\frac{r-r_h}{r_h}\right)^3\,
\\
\nonumber
l(r,\theta)&=&l_2(\theta)\left(\frac{r-r_h}{r_h}\right)^2 
 + O\left(\frac{r-r_h}{r_h}\right)^3. 
\end{eqnarray}
Since from general arguments the Hawking temperature $T$ is proportional
to the surface gravity
$\kappa $, $
T=\kappa /(2 \pi),
$
we obtain the relation
\begin{equation}
\label{temp}
T=\frac{f_2(\theta)}{2 \pi r_h \sqrt{q_2(\theta)} } 
\ . 
\end{equation}
One can show, with help of the $(r~\theta)$
Einstein equation which implies
\begin{eqnarray}
\label{con}
f_2  q_{2, \theta}=2q_2 f_{2,\theta},
\end{eqnarray}
that the temperature $T$, as given in (\ref{temp}), is indeed constant.

For the line element (\ref{metric}),
the  area $A$ 
of the event horizon  is given by 
\begin{equation}
\label{area}
A = 2 L \pi \int_0^\pi  d\theta \sin \theta
\frac{\sqrt{l_2(\theta) q_2(\theta)}}{f_2(\theta)} r_h^2.
\end{equation}
According to the usual thermodynamic arguments, the entropy $S$ is proportional 
to the area $A$  
\begin{equation}
\label{entropy}
S = \frac{A}{4G},  
\end{equation}
leading to the product
\begin{equation}
TS = \frac{r_h L}{4G} \int_0^\pi  d\theta \sin \theta
{\sqrt{l_2(\theta) }}. 
\end{equation}

The horizon parameter $r_h$ entering the boundary conditions
is not a physical parameter. We thus introduce the area parameter $x_{\Delta}$ with
\begin{equation}
x_{\Delta}=\sqrt{\frac{A}{4\pi L}}
\end{equation} 
to characterize the solutions.

Since the energy density of the matter fields is angle dependent at 
the horizon, the horizon will be deformed.
A suitable parameter to measure the deformation of the horizon
of the black string solutions is given by the ratio $\rho:=L_e/L_p$ of the
horizon circumference along the equator 
\begin{equation}
L_e= \left. \left(\int\limits_0^{2\pi} d \varphi \ e^{-a\psi/2}\sqrt{\frac{l}{f}} \
\sin\theta  \ r\right)\right|_{r=r_h,\theta=\pi/2} \ \ , 
\end{equation}
and along the poles
\begin{equation} 
L_p=2 \left. \left(
\int\limits_0^{\pi} d\theta  \  e^{-a\psi/2}\  \sqrt{\frac{q}{f}} \ r  
\right)\right|_{r=r_h, \varphi=0}  \ .
\end{equation}
Note that the non-uniform black string solutions \cite{Gubser:2001ac,Wiseman:2002zc} 
have also a deformed horizon, with a different physical origin, however.
\subsection{A computation of the Euclidean action}
The expression (\ref{entropy}) for the entropy can be derived in a more rigorous way by using
Euclidean quantum gravity arguments.
Here we start by constructing the path integral \cite{Gibbons:1976ue}
\begin{eqnarray}
\label{Z1}
Z=\int D[g]D[\chi]e ^{-iI[g,\chi]}
\end{eqnarray}
by integrating over all metric and matter fields between  some given initial and final
hypersurfaces, $\chi$ corresponding here to the SU(2) potentials.
By analytically continuing the time coordinate $t \to it$,
the path integral formally converges, and in the leading order one obtains
\begin{eqnarray}
\label{Z2}
Z \simeq e^{-I_{cl}}
\end{eqnarray}
where $I_{cl}$ is the classical action evaluated on the equations of motion
of the gravity/matter system.
In computing $I_{cl}$, one should supplement (\ref{action5})
with the boundary term
\begin{eqnarray}
\label{IGH}
I_{b}=-\frac{1}{8\pi G}\int_{\partial\mathcal{M}} d^{4}x\sqrt{-h}(K-K_0),
\end{eqnarray}
where $K$ is the trace 
of the extrinsic curvature for the boundary $\partial\mathcal{M}$ and $h$ is the induced 
metric of the boundary. 
In (\ref{IGH}) we have already subtracted the contribution of
the background spacetime which is taken to be flat space $M^4\times S^1$.

We note that the considered Lorentzian solutions of the EYM equations extremize also the
Euclidean action, $t \to it$ having no effects at the level of the equations of motion.
The value of $\beta$ is found here by demanding regularity 
of the Euclideanized manifold as $r \to r_h$,
which together with the expansion (\ref{expan-h}) 
gives $\beta=1/T$. The physical interpretation of this formalism is that
the class of regular stationary metrics forms an ensemble of
thermodynamic systems at equilibrium temperature
$T$ \cite{Mann:2002fg}.
$Z$ has  the interpretation of the partition function and we can
define the free energy of the system
$F=-\beta^{-1} \log Z$.
Therefore
\begin{eqnarray}
\label{i1}
\log Z=-\beta F=S-\beta M,
\end{eqnarray}
or
\begin{eqnarray}
\label{i2}
S=\beta M-I_{cl},
\end{eqnarray}
straightforwardly follows.

To compute $I_{cl}$, we make use of the Einstein equations, replacing the $R$ volume term with
$2R_t^t-16\pi G T_t^t$.
For our purely magnetic ansatz, the term $T_t^t$  exactly cancels the matter field
lagrangian in the bulk action
 $L_m=-1/2g^2Tr(F_{MN }F^{MN})$ (see also the general discussion in \cite{Visser:1993qa}).
The divergent contribution given by the surface integral term at infinity in $R_t^t$ is canceled by 
the boundary action (\ref{IGH}) and we arrive at the simple finite expression
\begin{eqnarray}
\label{itot}
I_{cl}=(2c_t-c_z)\frac{L\beta}{4\pi G}
-
\frac{\pi L }{2G}\int_0^\pi  d\theta \sin \theta
\frac{\sqrt{l_2(\theta) q_2(\theta)}}{f_2(\theta)} r_h^2
\end{eqnarray}
Replacing now in (\ref{i2}) (where $M$ is the mass-energy computed in Section 2.4), we find
\begin{eqnarray}
\label{i3}
S= \frac{\pi L}{2G}\int_0^\pi  d\theta \sin \theta
\frac{\sqrt{l_2(\theta) q_2(\theta)}}{f_2(\theta)} r_h^2,
\end{eqnarray}
which is one quarter of the event horizon area, as expected.

\section{Numerical solutions}
For a nonvanishing magnitude of the gauge potential
$A_5$ at infinity, $\eta$, it is convenient to work with dimensionless variables
by taking the rescalings $r \to r\eta g$ and $\Phi \to \Phi/\eta$. Thus the field
equations depend  only on the coupling
constants $\alpha=\sqrt{4\pi G}\eta$,
yielding the dimensionless mass 
(per unit length of the extra dimension)  $\mu= (4\pi G\eta^2)^{-1}M$,
the dimensionless tension being  $\sigma= (4\pi G\eta^2)^{-1}\tilde{\sigma}$.
For $\alpha=0$ (no gravity) and no dependence on the $z$-coordinate, 
the four dimensional picture
 corresponds to the SU(2)-YMH theory in a flat $M^4$ background. 
 
The solutions' properties crucially depend on the value of $m$ and
will be discussed separately for $m=0$ and $m=1$.
In each case, we start by presenting a review of the globally regular
vortex-type configurations which
turn out to be crucial in the understanding
of the domain of existence of the uniform black string solutions.

The spherically symmetric solutions are found by using the differential
equation solver COLSYS \cite{COLSYS}.
In the axially symmetric case, the resulting set of eleven partial
differential equations are solved by 
using the program FIDISOL, 
based on the iterative Newton-Raphson method. 
Details of the FIDISOL code are presented in \cite{FIDISOL}.
In this scheme, a new radial variable is introduced
which maps the semi infinite region $[r_h,\infty)$ to the closed region $[0,1]$.
Our choice for this transformation was $ x=1-r_h/r$.
Typical grids have sizes $100 \times 30$, 
covering the integration region 
$0\leq{x}\leq 1$ and $0\leq\theta\leq\pi/2$.
The typical  numerical error 
for the functions is estimated to be on the order of $10^{-3}$.

\subsection{$m=0$ vortex-type configurations} 
The equations of motion admit next to black string
solutions also vortex-type solutions which exist on the full
interval $[0:\infty)$.
The boundary conditions satisfied by these solutions as $r\to \infty$ 
and on the axes are similar to the black string case,
while as $r \to 0$ one imposes
\begin{eqnarray}
\label{r0MM}
H_{1}|_{r=0}=H_3|_{r=0}=0 \ , \ H_{2}|_{r=0}=H_4|_{r=0}=1 \ , 
\ \Phi_1|_{r=0}=\Phi_2|_{r=0}=0,
\\
\nonumber
\partial_r \psi|_{r=0}=\partial_r f|_{r=0}= \partial_r q|_{r=0}=
\partial_r l|_{r=0}= J|_{r=0}=0.
\end{eqnarray} 
%
\subsubsection{Spherically symmetric solutions} 
For $m=0,~n=1$, all metric and matter functions depend only on the radial coordinate $r$.
For the matter functions we then have 
\begin{equation}
\label{sph}
H_1=H_3= \Phi_2=0,~~H_2=H_4=K(r),~~\Phi_1=\Phi_1(r),
\end{equation}
while for the metric functions we get $J=0$ and $q=l$ 
such that the metric becomes diagonal:
\begin{equation}
\label{metricB}
ds^2=e^{-a\psi(r)}\left[-f(r)dt^2+\frac{q(r)}{f(r)}
\left(dr^2+r^2d\theta^2+r^2 \sin^2\theta d\varphi^2\right)\right]+
e^{2a\psi(r)} dz^2  \ .
\end{equation}
In \cite{Volkov:2001tb} it has been found that several branches of solutions 
can be constructed when the effective gravitational
coupling $\alpha$ is varied. While the first branch exists for 
$\alpha\in [0: 1.268]$, the successive branches
exist on smaller and smaller intervals of $\alpha$, e.g. the
 second branch  for $\alpha\in [0.312:1.268]$,
the third for $\alpha\in [0.312:0.419]$ and the fourth for 
$\alpha\in [0.395:0.419]$.
Further branches can be constructed in the interval  
$\alpha\in [0.395:0.419]$, but have not been
determined in detail. The endpoint of the branches
is reached when the interval in $\alpha$ shrinks to a 
point at $\alpha=\alpha_{cr}$. 
It has been noticed in \cite{Volkov:2001tb} that the gauge field
function $H_2(r)=H_4(r)\equiv K(r)$ becomes oscillating 
around a fixed point. The number of oscillations
increases along succesive branches and becomes infinite 
in the limit $\alpha\rightarrow \alpha_{cr}$.
 
For one fixed value of $\alpha$, we thus have one, two, three and four
globally regular vortex solutions for $\alpha\in [0:0.312[$, $\alpha\in ]0.419:1.268]$,
$\alpha\in [0.312:0.395[$ and $\alpha\in [0.395:0.419]$, respectively.
The mass of the fundamental solution on the first (the ``main'') branch
is lower than the mass of the solution on any successive 
branch for a fixed value of $\alpha$.
We would thus expect the ``higher'' solutions to be unstable with respect
to the fundamental solution.  

\subsubsection{Axially symmetric solutions} 
The spherically symmetric 
configurations admit axially symmetric generalisations, obtained by taking
$n>1$ in the matter ansatz.
A general discussion of these  solutions has been presented in \cite{Brihaye:2005pz}.
Here we report new results for $n=2$.
While in \cite{Brihaye:2005pz} only one branch of solutions in $\alpha$ has
been constructed, we have reconsidered this problem
and managed to construct a second branch of deformed non-abelian vortices. 
While the first (``main'') branch exists for $\alpha\in [0:1.28]$,
we constructed a second branch for $\alpha\in [0.88:0.1.28]$.
This is demonstrated in Figure 1, where we plot the mass of the solution
per winding number $\mu/2$
as function of $\alpha$. Clearly, the solutions on the second
branch have higher energy than those on the main branch for the same
value of $\alpha$. We also show the values of the
metric functions $f$ and $\psi$ at the origin (with $\theta=0$),
$f(0)$ and $\psi(0)$. Clearly, the solutions
on the two branches are different for one fixed value of $\alpha$.
$f(0)$ on the second branch of solutions is very close to zero.
This is also demonstrated in Figure 2, where
we compare the profiles of the metric functions $f$ and $\psi$
for $\alpha=0.8$ on the two branches.

We believe that several more branches exist, but were unable to construct them.
Again, the pattern of globally regular deformed vortex
solutions is important to understand the domain of existence of the
deformed black strings, which we will discuss below.

\subsection{$m=0$ black string configurations} 
\subsubsection{Spherically symmetric solutions} 
Here we present a detailed analysis of the model studied 
in \cite{Hartmann:2004tx,Brihaye:2005fz}.
In \cite{Hartmann:2004tx}, black string solutions for a fixed value of the
horizon value and varying gravitational coupling $\alpha$ have been constructed.
It has been found that the pattern of solutions is very similar to that
observed for non-abelian vortices. Several branches of solutions exist
and the extend of the branches in $\alpha$ gets smaller and smaller for successive branches.
In \cite{Brihaye:2005fz}, the gravitational coupling has been fixed to $\alpha=0.5$
and the horizon radius has been varied. Two branches of solutions in $x_{\Delta}$, which both exist
for $x_{\Delta}\in [0:x_{\Delta,max}]$ have been
found with $x_{\Delta,max}\approx 0.633$.
In the limit $x_{\Delta}\rightarrow 0$, the solutions on the first and second branch, respectively,
correspond to the globally regular non-abelian vortex solution on the main 
and second branch of vortex solutions for 
fixed $\alpha$. It has been suggested in \cite{Brihaye:2005fz} that two branches in $x_{\Delta}$ 
exist for all fixed values of $\alpha$. The numerical results presented here
confirm this expectation. In Figure 3 we show the domain of existence of
the black string solutions in the $(\alpha$-$x_{\Delta})$-plane.

As is clearly seen from this figure, the critical behaviour of the
solutions depends crucially on the choice of $\alpha$. For $\alpha\in [0:0.312]$
the solutions on the second branch tend to the Einstein-Maxwell-dilaton (EMD)
solutions for a finite value of $x_{\Delta}=x_{\Delta,cr}$. This is demonstrated in Figure 4
for $\alpha=0.2$, where the value of the gauge field function $H_2(x)=H_4(x)
\equiv K(x)$ at the horizon, $K(x_{\Delta})$, is shown as function of
$x_{\Delta}$. Clearly, the second branch of solutions ends at $K(x_{\Delta,cr})=0$,
which together with the boundary conditions at infinity tells us that
$K(x)\equiv 0$. 
 
In addition, the value of the gauge field function $\Phi_1$ at $x_{\Delta,cr}$ 
tends to one for $x_{\Delta}\rightarrow x_{\Delta,cr}$ (see Figure 5) 
such that $\Phi_1(x)\equiv 1$. The gauge field is thus trivial
and becomes that of an EMD solution.
At the same time, the metric function $\psi$, which in the 4-dimensional
picture corresponds to a dilaton, tends to the dilaton field of
the corresponding  EMD  solution. This can
e.g. be seen by comparison of the value of $\psi(x_{\Delta,cr})$ (see Figure 6)
with that of the EMD solution with same horizon value.
 
For $\alpha=0.5$, the picture is completely different. Again, two branches
of solutions exist, but now the second branch extends all 
the way back to $x_{\Delta}=0$. $K(x_{\Delta})$ and $\Phi_1(x_{\Delta})$
tend back to one, respectively zero, the values of the globally regular
vortex solution (see Figure 4 and Figure 5). The solutions
reach the globally regular vortex solutions discussed
in Section 3.2 for $x_{\Delta}\rightarrow 0$.
However, Figure 5 shows that the terminating solution
of the second branch is different from that of the first branch.
The limiting solutions of the first, respectively second branch
correspond to the fundamental and second globally regular vortex solution.
The reason why the second branch terminates in the second regular solution
relates to the fact that only two different globally regular vortex
solutions exist for $\alpha=0.5$. If we had chosen a value of $\alpha$ for
which more than two vortex solutions exist, the second branch of
black strings would terminate into the ``highest'' available solution,
e.g. for $\alpha\in [0.395:0.419]$, the second branch would terminate
in the 4th  globally regular solution. This behaviour
is demonstrated in Figure 3, where we indicate with ``2.'', ``3.''
``4.'' at which globally regular vortex solution the second branch
of black strings terminates.
 
\subsubsection{Axially symmetric solutions} 
We have constructed deformed black string solutions for $n=2$ for
three fixed values of $\alpha$. Our results are shown
in Figures~6-10, where we give the values of
the gauge field functions $H_2$, $\Phi_1$ and of the
metric functions $\psi$ and $J$ at the horizon as functions
of the area parameter $x_{\Delta}$. First, we remark that
-apart from the metric function $J$- 
the curves can hardly be distinguished for different
values of $\theta$. Very similar to what has been observed in the
case $n=1$, two branches of solutions exist. Again, the limiting
behaviour depends strongly on the choice of $\alpha$.
Since we don't have a detailed analysis of the branch structure
of the $n=2$ globally regular vortex-type configuration,
we cannot make a precise prediction 
to which solution the second branch tends. However, we can make qualitative
statements about the domain of existence of the deformed black strings
in the $(\alpha$-$x_{\Delta})$-plane.
We find that for $\alpha=0.1$, the second branch terminates
into an  EMD  solution with $H_2(x_{\Delta})$ tending
to zero at a finite $x_{\Delta}=x_{\Delta,cr}$ (see Figure 7), 
$\Phi_1(x_{\Delta})$ reaching one in this limit
(see Figure 8), while $\psi$ and $J$ take their respective
values of the corresponding EMD solution (see Figure 9 and
Figure 10). 
Note that the EMD is a
spherically symmetric solution, the curves for different $\theta$ should thus
meet at $x_{\Delta,cr}$. This is what we noticed in our computation.
For $\alpha=0.5$ and $\alpha=1.0$, the situation is different.
We find that the second branch of black string solutions reaches back 
to $x_{\Delta}=0$, where
it tends to a globally regular vortex-type solution. Since we don't know
the detailed branch structure of the globally regular $n=2$ solution,
we don't know which ``higher'' solution the limiting solution is.
Clearly, the limiting solution is axially symmetric, which can
be seen in Figure 10, where the curves for $j(x_{\Delta})$ (note that
$j$ is the transformed function given in Appendix B) 
ends at finite values for $x_{\Delta}=0$. 
If the solutions would be spherically symmetric the off-diagonal
component of the metric tensor would be vanishing, i.e. $j=J\equiv 0$, which
clearly is not the case here.

The domain of existence of the deformed black strings is qualitatively
very similar to that of the uniform black strings.
The values for $n=2$ are given in Table 1.

\begin{table}[t]
\caption{Maximal and critical value of $x_{\Delta}$ for deformed
black string solutions ($n=2$)}
\vspace{2mm}
\small
\begin{center}
\begin{tabular}{|c|c|c|c|}           
\hline
\raisebox{0mm}[4mm][2mm] {$\alpha$} & $x_{\Delta,max}$
& $x_{\Delta,cr}$ & Bifurcation of second branch with \\
\hline\hline
\multicolumn{1}{|l|}{\raisebox{0mm}[4mm]{$0.1$}} & $1.35$ & $0.88$ & EMD
\\[0.5mm]
\hline   
\multicolumn{1}{|l|}{\raisebox{0mm}[4mm]{$0.3$}} & $1.33$ & $0.72$ & EMD
\\[0.5mm]
\hline
\multicolumn{1}{|l|}{\raisebox{0mm}[4mm]{$0.5$}} & $1.30$ & 0.0 & regular
\\[0.5mm]
\hline
\multicolumn{1}{|l|}{\raisebox{0mm}[4mm]{$1.0$}} & $0.845$ & 0.0 & regular 
\\[0.5mm]
\hline
\end{tabular}
\vspace{-1mm}
\end{center}
\end{table}

We give the maximal horizon value $x_{\Delta,max}$ at which the two branches
meet as well as the critical horizon value $x_{\Delta,cr}$ where the second
branch of solutions ends. The last column indicates
at which solution the second branch terminates. Clearly for
$\alpha=0.1$ and $\alpha=0.3$, the second branch tends to the
EMD solution for $x_{\Delta}\rightarrow x_{\Delta,cr}$, while
for $\alpha=0.5$ and $\alpha=1.0$, the second branch extends all the way back
to $x_{\Delta}=0$, where it bifurcates with the globally regular solution. 
We note when comparing the results for the deformed black strings
with those for the uniform black strings that the former exist for much larger
values of the horizon value. 

In Figure 11, we show the ratio of the horizon circumference
along the equator $L_e$ and along the poles $L_p$, $\rho=L_e/L_p$
as function of $x_{\Delta}$ for three different values of $\alpha$.
This ratio is a direct measure for the deformation of the horizon
of the black string solution. For $\alpha=0.1$, $\rho$ stays very close
to one which, of course, is related to the weak gravitational coupling.
For both $\alpha=0.5$ and $\alpha=1.0$, the deformation of the horizon
is much stronger on the second branch as compared to the first branch.
Interestingly, the maximal deformation of the $\alpha=0.5$ solution
is larger than that of the $\alpha=1.0$ solution. This is likely related
to the fact that the $\alpha=0.5$ black string solutions have larger possible
horizon values than the $\alpha=1.0$ solutions.

In Figure 12, we show the temperature $T$ and the entropy $S$ of the
deformed black strings ($n=2$) as functions of $x_{\Delta}$.
We present the figure only for $\alpha = 0.5$ because
  the curves differ hardly for different $\alpha$ (e.g. $\alpha = 
0.2$, $\alpha = 1.0$). The numerics further indicates that, for fixed
  $x_{\Delta}$, the solution with the lower 
  mass possesses the lower  temperature. 
The temperature tends to infinity in the 
limit $x_{\Delta}\rightarrow 0$,
while the entropy tends to zero. 
This is not  surprising since as $x_{\Delta}\rightarrow 0$
the event horizon area tends also to zero, while the temperature of a
globally regular solution is arbitrary. 

The dependence of the mass of the solutions of the parameter  $x_{\Delta}$ is exhibited
in Figure 13 for two values of $\alpha$.
In Figure 14 we present a plot of the tension $\sigma$ of the solutions as a function of $x_{\Delta}$
for two values of $\alpha$. One can see that the behaviour of $\sigma$ resembles that of $\mu$, 
the tension
on the upper branch being higher that the corresponding value on the lower branch.
In Figure 15 the $(\mu, n_t)$ phase diagram is presented for $\alpha=0.1,~\alpha=0.5$.
This type of phase diagram was proven important in classifying the
non-uniform black strings and black holes on a cylinder \cite{Harmark:2003dg}.
One finds that these deformed black string solutions cover compact regions of the $(\mu,n_t)$ plane.
In particular there are solutions with $n_t > 1/2$, where $1/2$ is the vacuum value.

The energy density of the matter fields is angle dependent, and in particular
is not constant at the horizon.
The maximum of the energy density resides on the $\bar{\rho}$ 
axis, as seen in  Figure 16, 
where a three-dimensional plot of the $T_t^t$ 
component of the energy-momentum tensor is presented.

\subsection{$m=1$ black string configurations} 
A very different picture is found by taking $m=1$ in the asymptotic
boundary conditions, $i.e.$ $\Phi_1=\cos \theta$, $\Phi_2=\sin \theta$,
the asymptotics of $H_i$ being fixed by (\ref{asm1})
(here we consider the case $n=1$ only, although a number of $n=2$ configurations
have been also studied with similar qualitative results).
Different from $m=0$, no spherically symmetric solutions are found for this set
of boundary conditions.

Let us briefly recall the features of the corresponding vortex-type 
solutions discussed in \cite{Brihaye:2005pz}.
The boundary conditions at infinity
and at $\theta=0,\pi/2$ satisfied by these regular solutions are
similar to the black hole case.
The conditions (\ref{r0MM}) at $r=0$ are also fulfilled except 
for the $A_5$ potentials, which satisfy
$\cos \theta\, \partial_r\Phi_{1}-\sin \theta\,\partial_r
\Phi_{2}=0,~~
\sin \theta\, \Phi_{1}+\cos \theta\, \Phi_{2}=0$.

In the limit $\alpha \rightarrow 0$, 
a  branch of $m=1$ vortex-type solutions emerges from 
the uplifted version of the
$d=4$ flat spacetime MA
configurations in YMH theory \cite{Kleihaus:1999sx}.
 This branch ends at
a critical value $\alpha_{cr} \approx 0.65$.
Apart from this fundamental
branch, the $m=1$ solutions admit also
excited configurations, emerging in the $\alpha \to 0$ limit (after
a rescaling)  from the spherically symmetric solutions with $A_5=0$
(corresponding after dimensional reduction to solutions of a $d=4$
EYM-dilaton theory).
The lowest excited branch (the only case discussed in \cite{Brihaye:2005pz}), 
originating  from the one-node spherically
symmetric solution,
evolves smoothly from $\alpha=0$ to $\alpha_{cr}$ where it
bifurcates with the fundamental branch. 
The energy density
$\epsilon = -T_{t}^t$ possesses maxima at $z=\pm d/2$ and a saddle
point at the origin, and presents the typical form exhibited in the
literature on MA solutions \cite{Kleihaus:2000hx, Kleihaus:1999sx}.
The modulus of the fifth component of the gauge potential possesses
always two zeros at $\pm d/2$
on the $z-$symmetry axis.
 The excited solutions become infinitely
heavy as $\alpha \to 0$
while the distance $d$ tends to zero.

These regular solutions present black hole counterparts,
which in a four dimensional perspective, share many properties
with the corresponding EYMH black hole 
configurations with magnetic dipole hair \cite{Kleihaus:2000kv}.
For any fixed value of $\alpha$, 
$0 < \alpha < \alpha_{\rm cr}^{\rm reg}$,
we obtain two branches of black string solutions.
Imposing a regular event horizon at a small radius $x_{\Delta}$,
the lower branch of black string solutions emerges from the 
corresponding lower branch globally regular EYM vortex-type solution.
This branch of solutions  extends to a maximal value
$x_{\Delta,max}(\alpha)$.
Along this lower branch, the mass increases with increasing $x_{\Delta}$  
(see Figure 23).
Decreasing $x_{\Delta}$ from
$x_{\Delta,max}(\alpha)$, a second branch of solutions appears.
Along this upper branch  the mass decreases with decreasing $x_{\Delta}$,
reaching the regular upper branch MA  solution,
when $x_{\Delta} \rightarrow 0$.
We note that for the same event horizon radius, 
the mass of the upper branch solution is
higher.
In Figures 17-20 we illustrate the value of the functions $H_2$, $\Phi_1$, $\psi$ and $J$ 
at $x_{\Delta}$ for two values of $\alpha$. 
Similar to the $m=0$ case, these plots do not exhibit 
a strong dependence on the value of $\theta$ (except for $J(x_{\Delta})$).
The metric function $J(r,\theta)$ presents a nontrivial angular dependence, behaving
asymptotically as $J \sim J_0\sin^2 \theta/r$.
Other branches of solutions may exist as well, in particular
those emerging from 
multinode regular configurations.

In Figure 21, we show the ratio of the horizon circumference
along the equator $L_e$ and along the poles $L_p$, $\rho=L_e/L_p$
as function of $x_{\Delta}$ for  $\alpha=0.2$ and $\alpha=0.5$.
As seen in Figure 22, the Hawking temperature increases with decreasing $r_h$, diverging as
$r_h \to 0$.
While the entropy of the regular solutions is zero, it approaches a maximal value 
at $r_{h(max)}(\alpha)$.
Figure 24 presents a plot of the tension of the solutions as a function of $x_{\Delta}$
for two values of $\alpha$. As in the case of $m=0$ solutions,
the behaviour of $\sigma$ resembles that of $\mu$ and the tension
on the upper branch is higher than the corresponding value on the lower branch.
In Figure 25 the $(\mu, n_t)$ phase diagram is presented for $\alpha=0.2,~\alpha=0.5$.
We note again the existence of solutions with $n_t > 1/2$.

In Figure 26 we exhibit the energy density of a typical lower branch
 $m=1,~n=1$ solution, with $\alpha=0.2$, $r_h=0.04$.
 Note the different shape of $T_t^t$ as compared to the $m=0$ case, 
 with two extrema on the $z-$axis.

\subsection{$A_5=0$   solutions} 
It is interesting to consider 
the following consistent reduction of the ansatz (\ref{ansatz})
\begin{equation}
\Phi_1=\Phi_2=0,
\end{equation}
($i.e.$ no Higgs field will appear in the $d=4$ theory),
in which case one should also set ${\cal W}_{\mu}=0$.

The four dimensional picture one finds from (\ref{action4})
corresponds to a EYM-dilaton system
\begin{eqnarray}
\label{actionEYMD}
I_4=\int d^{4}x\sqrt{-\gamma }\Big[
\frac{1}{4\pi G}\big(
\frac{\mathcal{R} }{4}
-\frac{1}{2}\nabla_{\mu}\psi \nabla^{\mu}\psi \big)
-e^{2\psi/\sqrt{3}}\frac{1}{2g^2}Tr\{
{\cal F}_{\mu \nu }{\cal F}^{\mu \nu }\}
\Big],
\end{eqnarray} 
with a particular coupling between dilaton and gauge field.

Both particle like and black hole solutions of this system are known to exist.
Here we'll review their basic properties from a five dimensional perspective.
Although the configurations are again indexed by the set of two integers
$(m,~n)$, the $A_5=0$ EYM solutions present a number of distinct features.
First, it appears that 
there are no $m=0$ finite 
mass nonabelian vortices or black string solutions.
This can easily be proven for configurations 
which are spherically symmetric in four dimensions 
by applying the arguments in  \cite{bizon}. 
This implies that, without a Higgs field, there are no
nonabelian monopole solutions
in the EYMD system (\ref{actionEYMD}). 

However, nontrivial solutions are found by taking the $m=1$ set of boundary
conditions for the matter fields $H_i$.  First, 
for $n=1$, the four dimensional
solutions are again spherically symmetric being discussed  
in \cite{Lavrelashvili:1992ia}.
The corresponding $d=5$ vortices and black strings
are parametrized by the number $k$ of nodes of the gauge function
$K(r)$, the extremal abelian solution being approached as $k$ tends to infinity.
All these solutions turn out to be unstable in linearized perturbation
theory \cite{Lavrelashvili:1992ia}.

As found in \cite{Kleihaus:1997mn}, \cite{Kleihaus:1997ws},
these configurations admit axially symmetric generalisations, 
obtained for a winding number $n>1$.
Different from the $A_5 \neq 0$ case, 
here we  find  black string solutions
for any radius $r_h$ of the horizon.
These  deformed solutions are characterized by two integers,
the winding number $n$ 
and the node number $k$ of the purely magnetic gauge field.
The mass of these solutions increases with  $n,~k$.
With increasing node number the magnetically neutral black string solutions 
form sequences tending to limiting solutions with magnetic charge $n$,
corresponding  in a $d=4$ picture to EMD black hole solutions.
Although no proof exists in the literature,
we expect these configurations to be unstable, too.

A general  $(2m,n)$ set of $d=4$
EYM solutions  have been discovered recently in \cite{Ibadov:2005rb}, \cite{Ibadov:2004rt}.
There the gauge potentials $H_i$ satisfy a complicated $m$-dependent set of boundary conditons.
These solutions presumably admit dilatonic generalization within the theory (\ref{actionEYMD}),
describing in a  $d=5$ picture  new sets of nonabelian vortices and 
black strings.

\section{New  $d=4$ solutions from boosted $d=5$ configurations }
For vacuum solutions extremizing ({\ref{action4}),
it has been known for some time that, by taking the product of the
$d=4$ Schwarzschild
solution with a circle and
boosting it in the fifth direction, the entire family of electrically
charged
(magnetically neutral) KK  black holes is generated.

As remarked in \cite{Brihaye:2005pz}, a similar construction
can be applied to the solutions of the $d=4$ EYMH-U(1)-dilaton
theory (\ref{action4}).
Starting with a purely magnetic $d=4$ static configuration
$(\gamma_{\mu \nu},{\cal A},{\cal W},\Phi, \psi)$,
 and uplifting it  according to (\ref{metrica}), (\ref{SU2}), one finds
in this way a vortex-type (or black string) solution of the 
$d=5$ EYM theory.
The next step is to boost this solution in the 
 $(z,~t)$ plane
\begin{eqnarray}
\label{boost}
z=\cosh \beta~Z+\sinh \beta \tau,~~t=\sinh \beta~Z+\cosh \beta\tau.
\end{eqnarray}
The dimensional reduction of this EYM configuration
along the $Z-$direction provides a new 
solution in the $d=4$ EYMH-U(1)-dilaton theory.
For the specific ansatz considered in this paper one finds
\begin{eqnarray}
\label{new4D}
d \sigma^2&=&\bar{\gamma}_{\mu \nu}dx^{\mu}dx^{\nu}=
  e^{a(\psi-\bar{\psi})}\gamma_{tt}
  (d\tau-2\sinh \beta {\cal W}_{\varphi} 
d \varphi)^2+e^{a(\bar{\psi}-\psi)}d \ell^2 
\\
\nonumber
&=&
-\bar{f}(d\tau-2\sinh \beta {\cal W}_{\varphi} d \varphi)^2
+\frac{\bar{q}}{\bar{f}}
(d r^2+ r^2 d \theta^2 )
+  \frac{\bar{l}}{\bar{f}} r^2 \sin ^2 \theta d\varphi^2,
\end{eqnarray}
the new  functions being expressed in terms of the initial solution as
\begin{eqnarray}
\label{new-mf}
\bar{\psi}&=&\psi+\frac{1}{2a}\log (\cosh^2 \beta
-e^{-3a\psi}f \sinh^2 \beta),
\\
\bar{f}&=&f\sqrt{\cosh^2 \beta-e^{-3a\psi}f \sinh^2 \beta},~~\bar{q}=q,~~
\bar{l}=l,
\end{eqnarray}
for the dilaton and metric functions, while the expression of the new gauge potentials is
\begin{eqnarray}
\label{new-gauge}
\nonumber
&&\bar{A_r}=A_r,~~\bar{A_\theta}=A_\theta,~~
\bar{A_\varphi}=A_\varphi-2\Phi {\cal W}_{\varphi}
\frac{e^{-2a\psi}f\sinh^2 \beta}{e^{2a \psi}\cosh^2 \beta-e^{-a\psi}f \sinh^2 \beta}~,
\\
&&\bar{A_\tau}=\Phi \sinh \beta 
\frac{e^{-a\psi}f}{e^{2a \psi}\cosh^2 \beta-e^{-a\psi}f \sinh^2 \beta}~,
\\
\nonumber
&&\bar{{\cal W}}_{\varphi}=\frac{e^{2a\psi}\cosh \beta~{\cal
W}_{\varphi}}
{e^{2a\psi}\cosh^2 \beta-e^{-a\psi}f\sinh^2 \beta}~,
~~
\bar{{\cal W}}_{\tau}=\frac{1}{2}
\frac{(e^{2a\psi}-e^{-a\psi}f)\sinh \beta \cosh \beta}
{e^{2a\psi}\cosh^2 \beta-e^{-a\psi}f\sinh^2 \beta}.
\end{eqnarray}
The new Higgs is
\begin{eqnarray}
\label{new-Higgs}
\bar{\Phi}=\Phi \cosh \beta.
\end{eqnarray}
To support a boosting given by a parameter $\beta$, one finds the condition  
\begin{eqnarray}
\label{tr3}
e^{-3a\psi}f\tanh^2<1
\end{eqnarray}
which turns out to be satisfied by all considered configurations.

For axially symmetric solutions, the new four dimensional line 
element presents a nonzero extradiagonal
metric component $\bar{\gamma}_{\varphi \tau}$, and thus describes a rotating spacetime.
Also, the electric potentials $ \bar{{\cal A}}_\tau$ of the $d=4$ 
SU(2) field is nonzero, being proportional to the Higgs field.

This procedure applied to globally regular MM and MA configurations generates
charged rotating solutions.
As discussed in \cite{Brihaye:2005pz},
although they will rotate locally, the total angular
momentum of the MM solutions is zero, and the spacetime consists in two
regions rotating in opposite directions. The solutions with a 
zero net magnetic charge possess a nonvanishing angular momentum proportional to the 
magnetic charge 
(the rotating solutions found recently in EYMH theory  
present the same qualitative picture \cite{Kleihaus:2005fs}).
In this construction, 
the causal structure is not affected by the
boosting procedure, $i.e.$  $\bar{\gamma}_{\varphi \varphi}>0$,  
(no closed timelike curves) and no event horizon occurs in the new solutions
(here we take also $-\infty<\tau<\infty$ and ignore the causal problems implied
by performing the transformation
(\ref{boost}) with an extra-$S^1$  direction).

The same procedure applied to static, axially symmetric black hole configurations
gives rotating black hole solutions.
For the new line element (\ref{new4D}),
the event horizon is a Killing horizon of the Killing vector
$\xi=\partial/\partial_{\tau}+\Omega_H\partial/\partial_{\varphi}$ \cite{Wald:rg}.
Here $\Omega_H=- \bar{\gamma}_{\tau\varphi}/\bar{\gamma}_{\varphi\varphi}$ 
(evaluated at the event horizon)
corresponds to the event horizon velocity.
The resulting solutions have a number of interesting properties.
For the specific ansatz used in (\ref{new4D}), one can see that
the event horizon location is unafected by the boosting procedure $\xi_{\mu}\xi^{\mu}(r_h)=0$,
while $\Omega_H=0$, $i.e.$ the event horizon is not rotating with respect 
to infinity.
Therefore 
the causal structure of the initial
solution is unchanged by the generation procedure, and no ergoregion is found, since 
$\bar{f}>0$ outside the event horizon. 
Moreover, similar to the regular case,
the ADM angular momentum of the black hole
solutions with a nonavanishing magnetic charge is zero, although
they rotate locally,  while
the $m=1$ solutions have a nonzero angular momentum.

A detailed discussion of the properties of the
$d=4$ EYMH-U(1)-dilaton rotating solutions
generated by boosting static axially symmetric configurations
in the  $x^5$-direction 
will be given elsewhere.
\subsection{$d=4$ spherically symmetric dyonic black holes}
For the rest of this Section we'll concentrate on the simpler 
spherically symmetric case.
The black string solutions of this model are discussed in 
\cite{Hartmann:2004tx}.
The four dimensional initial picture is straightforward, and consists
in magnetic monopole black holes in EYMH-dilaton theory ($i.e.$ no U(1) field
${\cal W}_{\mu}=0$).
After boosting, the new $d=4$ solutions  correspond to 
 black hole dyons, which, different from other cases 
discussed in the literature have also a nonvanishing $U(1)$ electric potential
(the spherically symmetric,
globally regular counterparts of these solutions have been constructed in
\cite{Brihaye:2004kh} by directly solving the field equations).

For the particular case of spherical symmetry, the 
relations (\ref{new4D}) read 
\begin{equation}
\label{metric4d}
d\sigma^2= -\bar{f}(r)d\tau^2
+\frac{\bar{q}(r)}{\bar{f}(r)}\left(dr^2+r^2d\theta^2+r^2 \sin^2\theta d\varphi^2\right),
\end{equation}
the new metric functions being determined by (\ref{new-mf}).
The four dimensional YM fields are  
\begin{eqnarray}
\label{A5}
\bar{A}_r=0,~~\bar{A}_{\theta}=(1-K(r))\tau_{\varphi}^{1 },
~~\bar{A}_{\varphi} =-(1-K(r))\sin \theta\tau_{\theta}^{1 },
~~
\bar{A}_{\tau} =\sinh \beta~\Phi_1(r)~\tau_r^1,
\end{eqnarray}
$K(r)$ being the magnetic nonabelian potential of the initial monopole solution.
The transformed Higgs field of the four dimensional theory is $H=\cosh \beta~\Phi_1$,
while the new dilaton $\bar{\psi}$ is given by
\begin{equation}
\label{dil-news}
e^{a \bar{\psi}}=
 e^{a\psi(r)}\sqrt{\cosh^2 \beta -e^{-3 a \psi(r)} f \sinh^2 \beta}.
\end{equation}
One can see that, similar to the solutions of the  (E-)YMH theory, 
the magnitude at infinity of the electric potential 
is restricted to be less than that of the Higgs field.
The Maxwell field which appears as a result of the boosting procedure
 possesses a nonvanishing electric potential $\bar{{\cal W}}_{\tau}$
which can be read from (\ref{new-gauge}).

The location of the event horizon is unaffected by boosting, while the relation 
between the  Hawking temperature and entropy
of the dyonic black holes and the corresponding quantities of the monopole solutions 
is $\bar{T}=T/\cosh \beta$, $\bar{S}=S\cosh \beta$.
The properties of a dyon solution can be predicted from the ``seed''
 configuration, in particular the domain of existence in the 
$(\alpha,r_h)$ plane.
The magnetic potential $K(r)$ vanishes at infinity
which gives a unit magnetic charge, while the nonabelian electric charge 
determined as $Q_e^{(n)}=\lim_{r\rightarrow \infty} r^2 \partial_r \bar{A}_{\tau}$
is
\begin{eqnarray}
Q_e^{(n)}=-\frac{\sinh\beta}{2}\left(3ad-(2h_1+M)+(3ad-2M)\cosh 2\beta\right) ,
\end{eqnarray}
(where $h_1=\lim_{r\rightarrow \infty} r^2 \partial_r \Phi$).
The  ADM mass and the abelian charge  of these solutions
are determined in terms of the mass $M$ and dilaton charge $d$ of the initial configurations
\begin{eqnarray}
\bar{M}=M(\cosh^2 \beta-\frac{1}{2}\sinh^2 \beta)-\frac{3}{4}a d \sinh^2 \beta,
~~
Q_e^{(a)}=\frac{1}{4}(3ad-2M)\sinh 2 \beta ,
\end{eqnarray}
while the new dilaton charge (defined as 
$\bar{d}=\lim_{r\rightarrow \infty} r^2 \partial_r \bar{\psi}$) is
\begin{eqnarray}
\bar{d}=d (\cosh^2 \beta+\frac{1}{2} \sinh^2\beta)-\frac{1}{a} M \sinh^2 \beta.
\end{eqnarray}
Other properties of these solutions can easily be deduced from the analysis presented 
in Section 3.2.1.

\section{Conclusions}
In this paper we have considered black string solutions in
the $d=5$ SU(2) EYM theory. From a four dimensional perpective,
these solutions correspond to
spherically and axially symmetric  
black holes sitting inside the center of
(multi)monopoles and monopole-antimonopole pairs.

For a class of regular configurations corresponding to 
solutions with a net magnetic charge in the $d=4$ theory,
we have presented new results
for a second branch of solutions. 
The domain of existence
in the $(\alpha$-$x_{\Delta})$-plane
of the black string counterparts of these 
configurations has been determined.
We find that when the gravitational coupling $\alpha$ is fixed,
two branches of solutions exist. The second branch terminates
into Einstein-Maxwell-dilaton solutions for values of the
gravitational coupling for which only one globally regular vortex solution
exists, respectively into the ``highest'' available globally regular
vortex solution for values of $\alpha$ where more than one vortex solution
exists.
We have also presented numerical arguments for the existence of a different type of 
$d=5$ EYM black strings, corresponding in a four dimensional picture to
black holes
located in between a monopole-antimonopole pair.

In this context,
we have proposed a simple procedure to generate new
$d=4$ electrically charged solutions
with nonabelian matter fields
starting with static $d=5$ EYM black string solutions.

In Appendix A an argument has been presented against the existence
of hyperspherically symmetric black hole solution 
with resonable asymptotics in SU(2) Einstein-Yang-Mills
theory in $d=5$. 
Therefore, the solutions discussed in this paper as well the regular counterparts
represent the simplest nontrivial configurations in $d=5$ EYM theory,
they providing also another couterexample to the no hair 
conjecture in five dimensions.

Concerning the stability of these nonabelian black holes, we
expect the $m=0$ solutions with a 
nonzero magnetic charge to be stable in a certain region 
of the parameter space. However,
all $m=1$ configurations are presumably unstable, like their 
MA-flat space counterparts \cite{Kleihaus:1999sx}.

One may  speculate about the existence of 
nonuniform vortices and black strings with nonabelian matter, with
a dependence on the extra $z$-coordinate.
Similar to the vacuum case, we expect these solutions
to emerge from the uniform EYM configurations
for a critical value of the mass.
However, the five dimensional gravity presents also black ring solutions,
with an horizon topology $S^2 \times S^1$, which approaches at infinity
the flat $M^5$ background. 
A vacuum black ring can be constructed in a heuristic way by taking the neutral black string,
bending the extra dimension and spinning it along the
circle direction just enough
so that the gravitational attraction 
is balanced by the centrifugal force.
This is a neutral rotating ring, obtained in Ref. \cite{Emparan:2001wn}
as a solution of the $d=5$ vacuum Einstein equations.
Generalizations of this solution for an abelian matter content are known in the literature.
Nonabelian versions of these configurations are also likely to exist, 
and will necessarily have an electric field.
However, the construction of such solutions represents a
difficult challenge. 
\\
\\
\\
\\
{\bf\large Acknowledgements} \\
YB is grateful to the
Belgian FNRS for financial support. BH thanks the University of Mons, where
part of this work was done, for hospitality.
The work of ER is carried out
in the framework of Enterprise--Ireland Basic Science Research
Project
SC/2003/390 of Enterprise-Ireland.
%
%
%
%
%

\newpage
\textbf{\Large Appendix A: No $d=5$ spherically symmetric black hole solutions}
\\
\\
Following the notations in \cite{Okuyama:2002mh},
we consider a static, spherically symmetric five-dimensional spacetime, with a metric 
given in Schwarzschild coordinates  by 
\begin{eqnarray}
ds^2= -f(r) e^{-2\delta(r)}dt^2+\frac{dr^2}{f(r)} +r^2
( d\psi ^2 + \sin ^2\psi \left(d\theta ^2+\sin^2 \theta ~d\varphi ^2) \right),
\end{eqnarray}
where
\begin{eqnarray}
f(r)&=&1-\frac{\mu(r)}{r^2}.
\end{eqnarray}
For black hole solutions $f(r_h)=0$ for some $r_h>0$, while 
$f'(r_h)>0$ and $\delta(r_h)$ stays finite.

The construction of a sherically symmetric Yang-Mills ansatz for a gauge group SU(2)
has been carefully discussed in \cite{Okuyama:2002mh}.
The expression of the nonabelian connection in this case is
\begin{eqnarray}
A_r^a&=&0,~~
A_\psi^a=(0, 0, w),~~
A_\theta^a=(w\sin\psi, -\cos\psi, 0),
\\
\nonumber
A_\varphi^a&=&(\cos\psi\sin\theta, w\sin\psi\sin\theta, -\cos\theta),~~A_t^a=0,
\end{eqnarray}
in terms of only one function $w(r)$.
With the above ansatz, we find the Einstein-Yang-Mills equations
\begin{eqnarray}
\label{eqs}
\mu'=2r\left[fw'^2+\frac{(1-w^2)^2}{r^2}\right],
~~\delta'=-\frac{2}{r}w'^2,
\\
\nonumber
\frac{1}{r}(rfe^{-\delta}w')'+\frac{2}{r^2}e^{-\delta}w(1-w^2)=0. 
\end{eqnarray}
Introducing a new variable 
\begin{eqnarray}
z=2 \ln r
\end{eqnarray}
 the equations (\ref{eqs}) imply the relations
\begin{eqnarray}
&&{d\mu \over dz}=4f\left({dw\over dz}\right)^2
+\left(1-w^2\right)^2
\label{basic_eqn1}\\
&&f{d^2w\over dz^2}+\left[e^{-z}\mu  -e^{-z} (1-w^2)^2\right]{dw\over dz}
+{1\over 2}w(1-w^2)=0,  
\label{basic_eqn2}
\end{eqnarray}
with
$
f=1-e^{-z}\mu ,
$
the function $\delta$ being eliminated.
To find a proof for the nonexistence 
of finite mass solutions of the above system, following \cite{Okuyama:2002mh},
it is convenient
to introduce the function
\begin{eqnarray}
E={1\over 2}f \left({dw \over dz}\right)^2-{1\over 8}\left(1-w^2\right)^2,
\end{eqnarray}
satisfying the equation
\begin{eqnarray}
{dE\over dz}=-4e^{-z}\left({dw \over dz}\right)^2
\left(E+{\mu\over 8}\right).
\label{eqE}
\end{eqnarray}
It is obvious that $E(r_h)<0$; at the same time, as proven in \cite{Okuyama:2002mh}, 
$E\to +0$ as $r \to \infty$, for finite mass solutions.
Therefore, if the solution is regular everywhere, $E$ must vanish at some
finite point $z_0$, and $dE/dr \geq 0$ there (when there are several positive roots of $E$,
we take the largest one).
However, another point should exist $z_1>z_0$ such that  $dE/dz = 0$ $i.e.$ the function
$E$ should present a positive maximum for some value of $z$.
Now we integrate the equation (\ref{eqE}) between $z_0$ and $z_1$ and find
\begin{eqnarray}
E(z_1)=-4\int_{r_0}^{r_1}e^{-z}\left({dw \over dz}\right)^2
\left(E+{\mu\over 8}\right) dz<0,
\end{eqnarray}
which contradicts $E(z_1)>0$. Therefore $E(z)$ should vanish identically and 
one finds no $d=5$ finite mass, spherically symmetric EYM configurations.
Note that this argument does not exclude the existence of configuration with
a diverging mass functions as $r \to \infty$.
In fact, such solutions can easily be found
and share many properties with the 
configurations without an event horizon discussed in \cite{Volkov:2001tb}. 
\\
\\
\\
\textbf{\Large Appendix B: Relation to a previous EYM string ansatz}
\\
\\
The $m=0$ deformed black strings discussed in  Ref. \cite{Brihaye:2005fz} have been found for
a slightly different ansatz.
While the ansatz (\ref{metrica}), (\ref{SU2}) admits
a straightforward KK picture and is useful
for the determination of the corresponding
rotating solutions, the notation of \cite{Brihaye:2005fz} turned out to be
more convinient for numerical computation of $m=0$ configurations. 

For completeness, we present here the relation between these two ans\"atze.
The five-dimensional metric parametrization used in \cite{Brihaye:2005fz}  is
\begin{equation}
ds^2=
e^{-\xi}\left[-\tilde{f}dt^2+\frac{\tilde{m}}{\tilde{f}}(dr^2+ r^2d\theta^2)+
\frac{\tilde{l}}{\tilde{f}} r^2 \sin^2\theta\left(d\varphi+j dz\right)^2  \right]
+e^{2\xi} dz^2 
\ , \end{equation}
while the YM matter ansatz corresponds to
\begin{eqnarray}
\label{ansatz1}
A_{\mu}dx^{\mu}&=&
\frac{1}{2g r}\left[\tau_{\varphi}^{n}
\left(\tilde{H}_1dr+(1-\tilde{H}_2)rd\theta\right)-n\left
(\tau_r^n \tilde{H}_3 +\tau_{\theta}^n (1-\tilde{H}_4)\right) r\sin\theta d\varphi \right.
\nonumber \\
&+& \left. \left(\tilde{\Phi}_1\tau_r^n+\tilde{\Phi}_2\tau_{\theta}^n\right)dz \right] 
\,
\end{eqnarray}
where $\tilde{f}$, $\tilde{l}$, $\tilde{m}$, $\xi$, $j$, $\tilde{H}_i$, $=1,2,3,4$ 
and $\tilde{\Phi}_1$, $\tilde{\Phi}_2$ are again functions of $r$ and $\theta$ only. 
They are related
to the ``untilded'' functions appearing in (\ref{metrica}), (\ref{SU2}) as follows:
\begin{eqnarray}
e^{a \psi}&=&
e^{\xi}\left(1+e^{-3\xi}\frac{\tilde{l}}{\tilde{f}} j^2r^2 \sin^2\theta\right)^{1/2} \ \ , \ \
J=  \frac{j}{2} 
\frac{\tilde{l}}{\tilde{f}} e^{-3\xi} r^2 \sin^2\theta \left(1+e^{-3\xi}\frac{\tilde{l}}{\tilde{f}} 
j^2 r^2 \sin^2\theta\right)^{-1} \ , \nonumber \\
f&=&\tilde{f} \left(1+e^{-3\xi}\frac{\tilde{l}}
{\tilde{f}} j^2 r^2 \sin^2\theta\right)^{1/2} \ \ , \ \ 
m=
\tilde{m} \left(1+e^{-3\xi}\frac{\tilde{l}}{\tilde{f}}j^2 r^2 \sin^2\theta\right) \ \ , \ \ 
l= \tilde{l} \ 
\end{eqnarray}
for the metric functions and
\begin{eqnarray}
H_1=\tilde{H}_1 \ , \ \ H_2=\tilde{H}_2\ , \ \ 
H_3=\tilde{H}_3+\frac{2g}{n} \frac{J \Phi_1}{\sin\theta} \ , \ \ 
H_4=\tilde{H}_4-\frac{2g}{n}  \frac{J \Phi_2}{\sin\theta} 
\ , ~~~\Phi_1=\tilde{\Phi}_1 \ , ~~~\Phi_2=\tilde{\Phi}_2  
\end{eqnarray}
for the gauge fields functions.
Of course, the final results are the same for both ans\"atze.

  
\newpage
\setlength{\unitlength}{1cm}

\begin{picture}(16,16)
\centering
\put(0,0){\epsfig{file=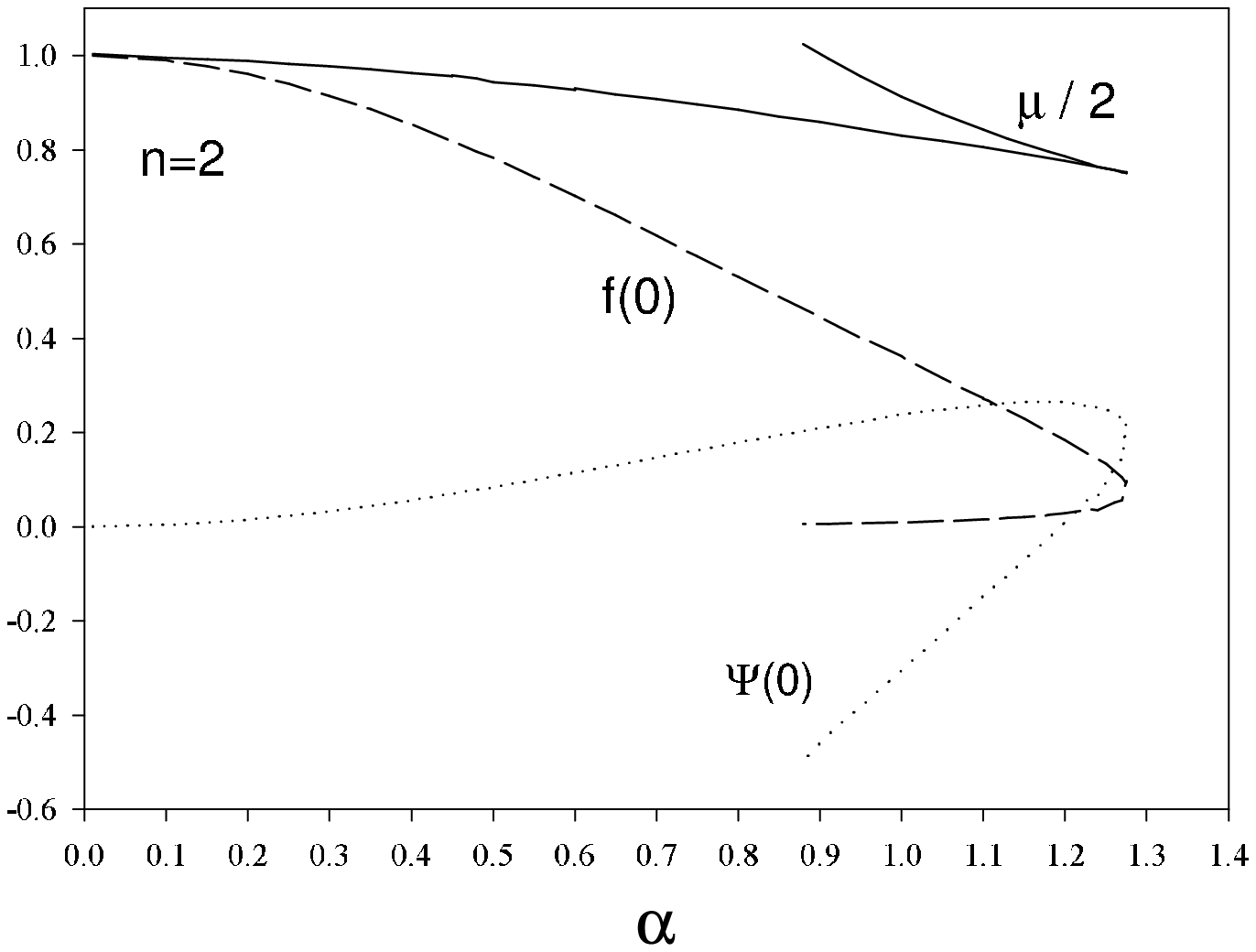,width=16cm}}
\end{picture}
\\
\\
\textbf{Figure 1.} The value of the metric functions $f$, $ \psi$ 
at the origin (with $\theta=0$),
$f(0)$, $ \psi(0)$ are shown as function of $\alpha$ for the deformed
non-abelian vortices with $m=0,~n=2$. The upper and lower curve
corresponds to the 1., respectively 2. branch of solutions. 
Also shown is the mass per winding
number of the solutions, $\mu/2$, 
where the 1. branch has lower energy than the 2.branch.

\newpage

\setlength{\unitlength}{1cm}

\begin{picture}(16,16)
\centering
\put(0,0){\epsfig{file=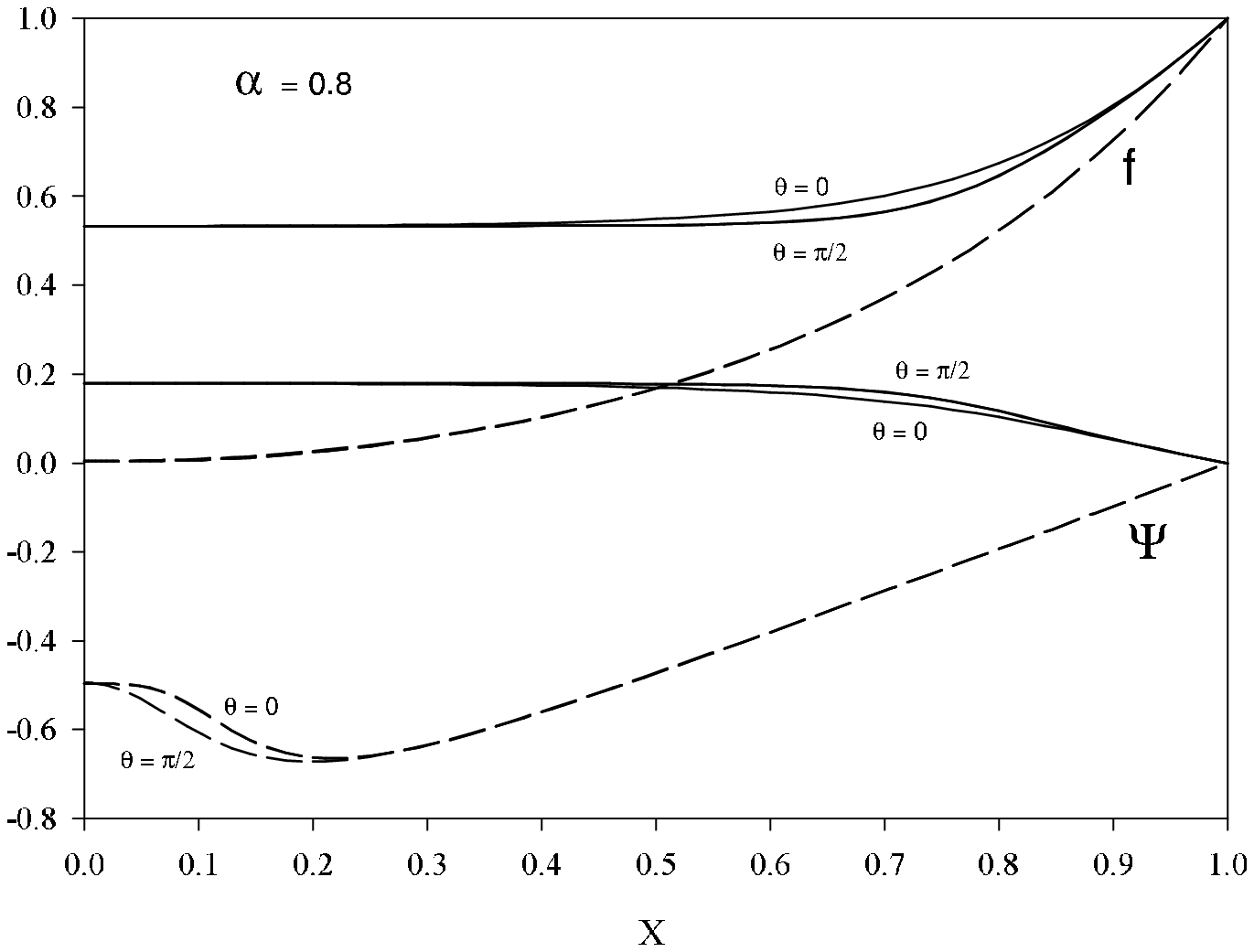,width=16cm}}
\end{picture}
\\
\\
\textbf{Figure 2.} The profiles of the metric functions $f$ and $\psi$ are shown
for $\alpha=0.8$ on the first (``main'') branch (solid) and on the second
branch (dashed) of $m=0$ non-abelian deformed vortex solutions ($n=2$).
$x$ here denotes the compactified coordinate $x\equiv r/(1+r)$.
Note that the two solid curves close to each other distinguish between
$\theta=0$ and $\theta=\pi/2$.
\\
\\

\newpage
\setlength{\unitlength}{1cm}

\begin{picture}(16,16)
\centering
\put(0,0){\epsfig{file=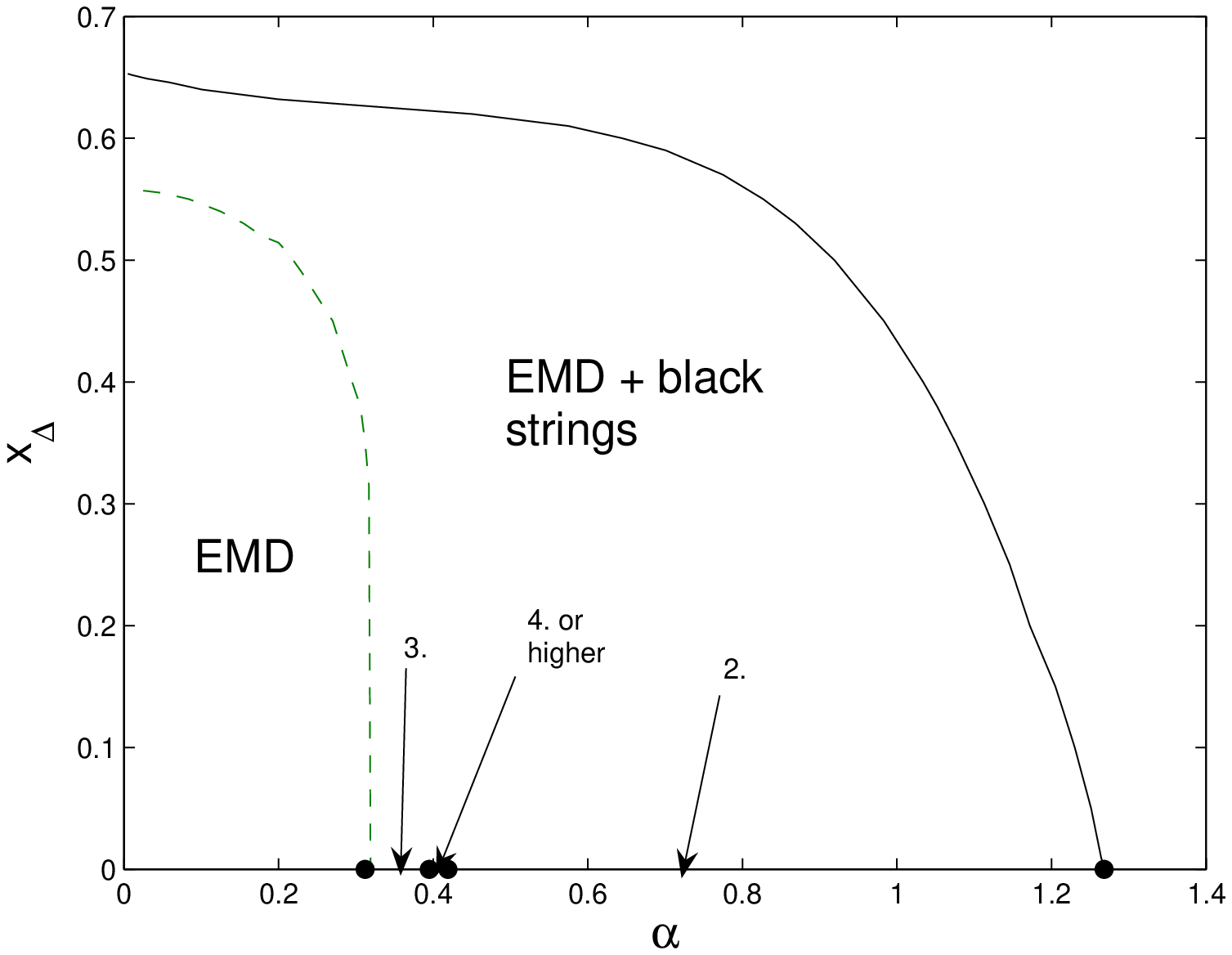,width=16cm}}
\end{picture}
\\
\\
\textbf{Figure 3.} 
The domain of existence of the $m=0$ uniform black string
solutions in the $(\alpha$-$x_{\Delta})$-plane is shown. ``2.'', ``3.'' and ``4. or higher''
indicate to which non-abelian vortex solution the black strings on the second
$x_{\Delta}$-branch tend in the limit $x_{\Delta}\rightarrow 0$.
\newpage
\setlength{\unitlength}{1cm}

\begin{picture}(16,16)
\centering
\put(0,0){\epsfig{file=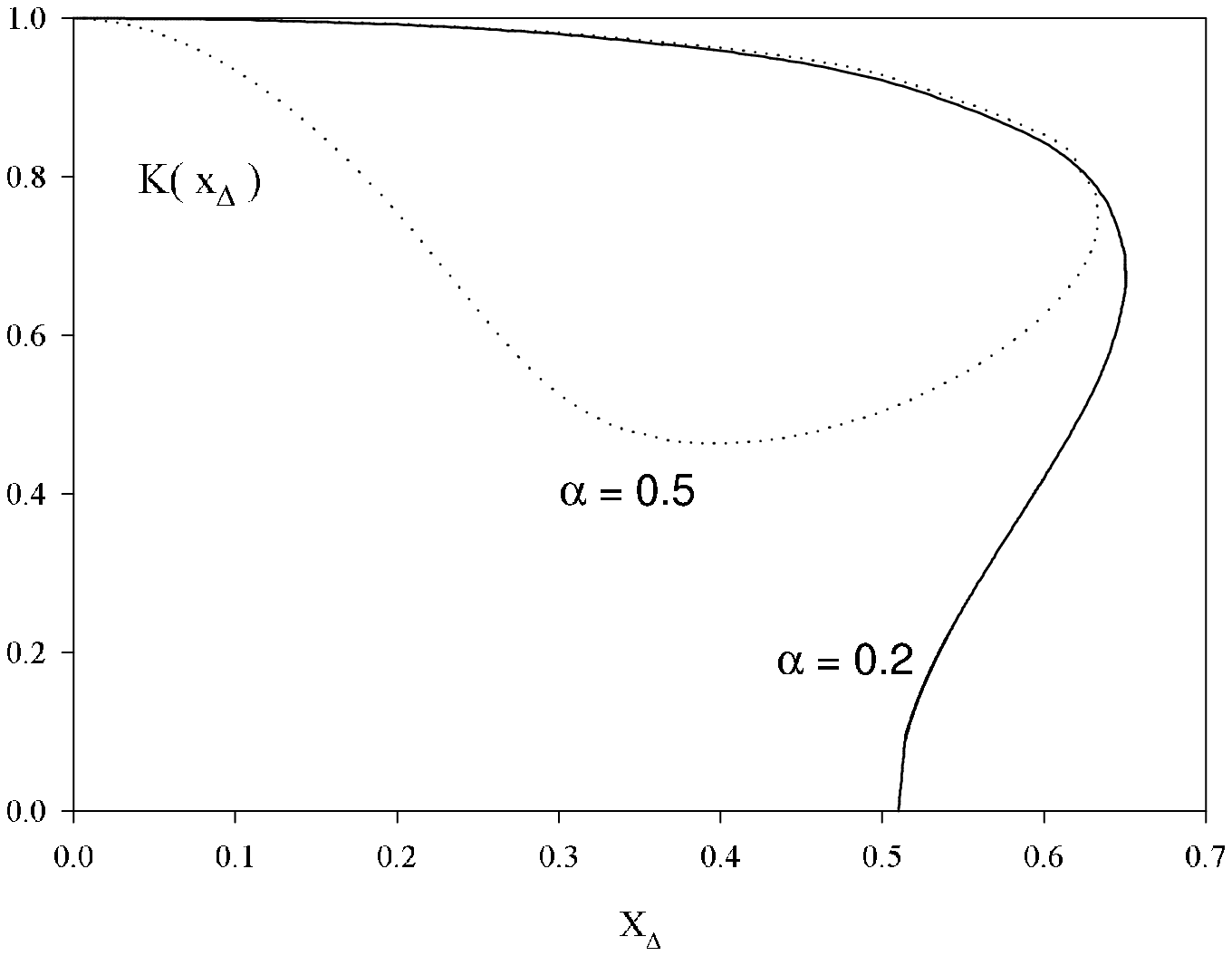,width=16cm}}
\end{picture}
\\
\\
\textbf{Figure 4.} The value of the gauge field function of the uniform
black string ($m=0,~n=1$) at the horizon $H_2(x_{\Delta})
=H_4(x_{\Delta})\equiv K(x_{\Delta})$  is
shown as function of $x_{\Delta}$ for $\alpha=0.2$ and $\alpha=0.5$.
The upper and lower curves correspond to the 1., respectively 2. branch of solutions. 
\\
\\

\newpage
\setlength{\unitlength}{1cm}

\begin{picture}(16,16)
\centering
\put(0,0){\epsfig{file=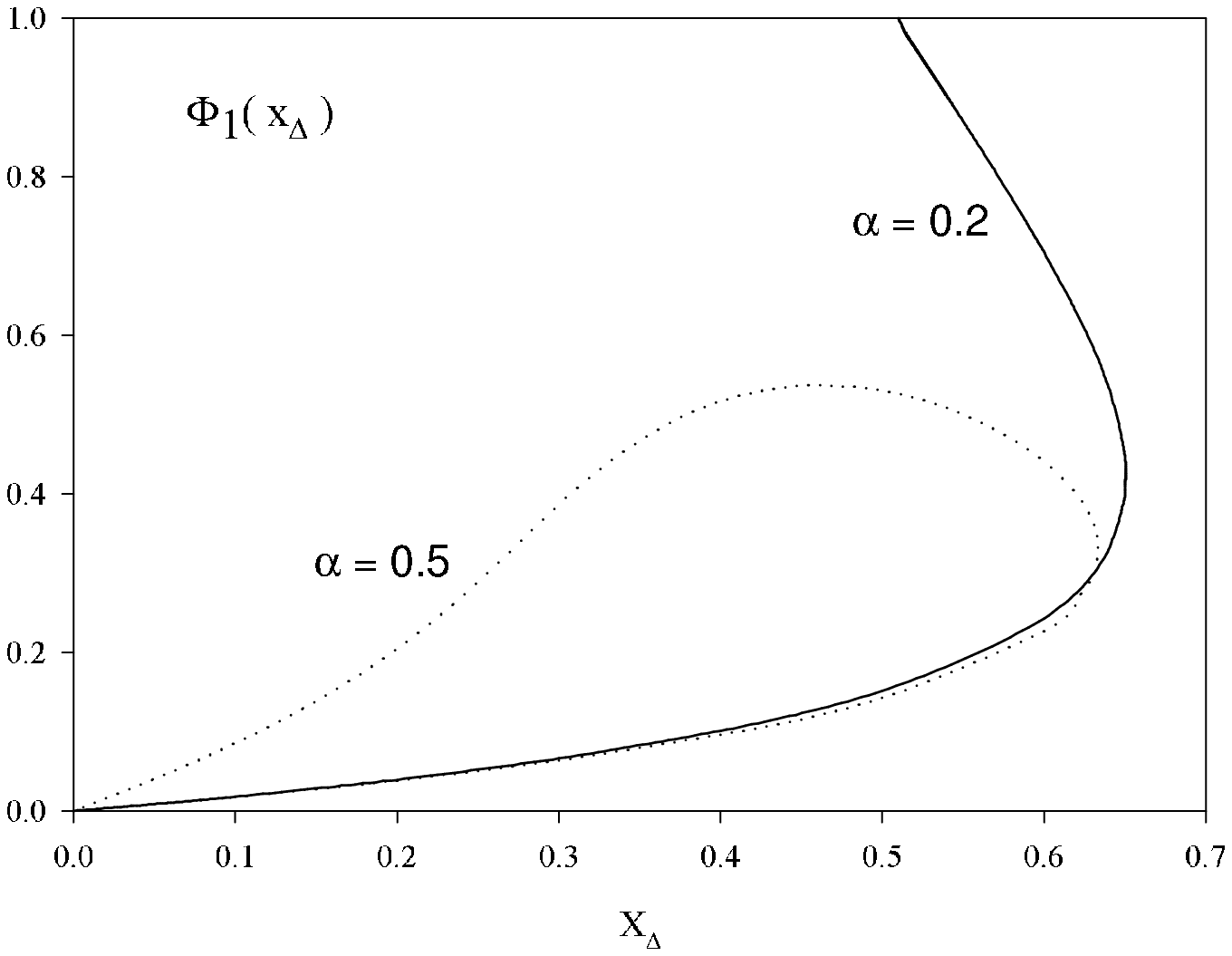,width=16cm}}
\end{picture}
\\
\\
\textbf{Figure 5.} The value of the gauge field function $\Phi_1$ of the uniform black string
($m=0,~n=1$) at the horizon, $\Phi_1(x_{\Delta})$, is
shown as function of $x_{\Delta}$ for $\alpha=0.2$ and $\alpha=0.5$.
The lower and upper curves correspond to the 1., respectively 2. branch of solutions.  
\\
\\

\newpage
\setlength{\unitlength}{1cm}

\begin{picture}(16,16)
\centering
\put(0,0){\epsfig{file=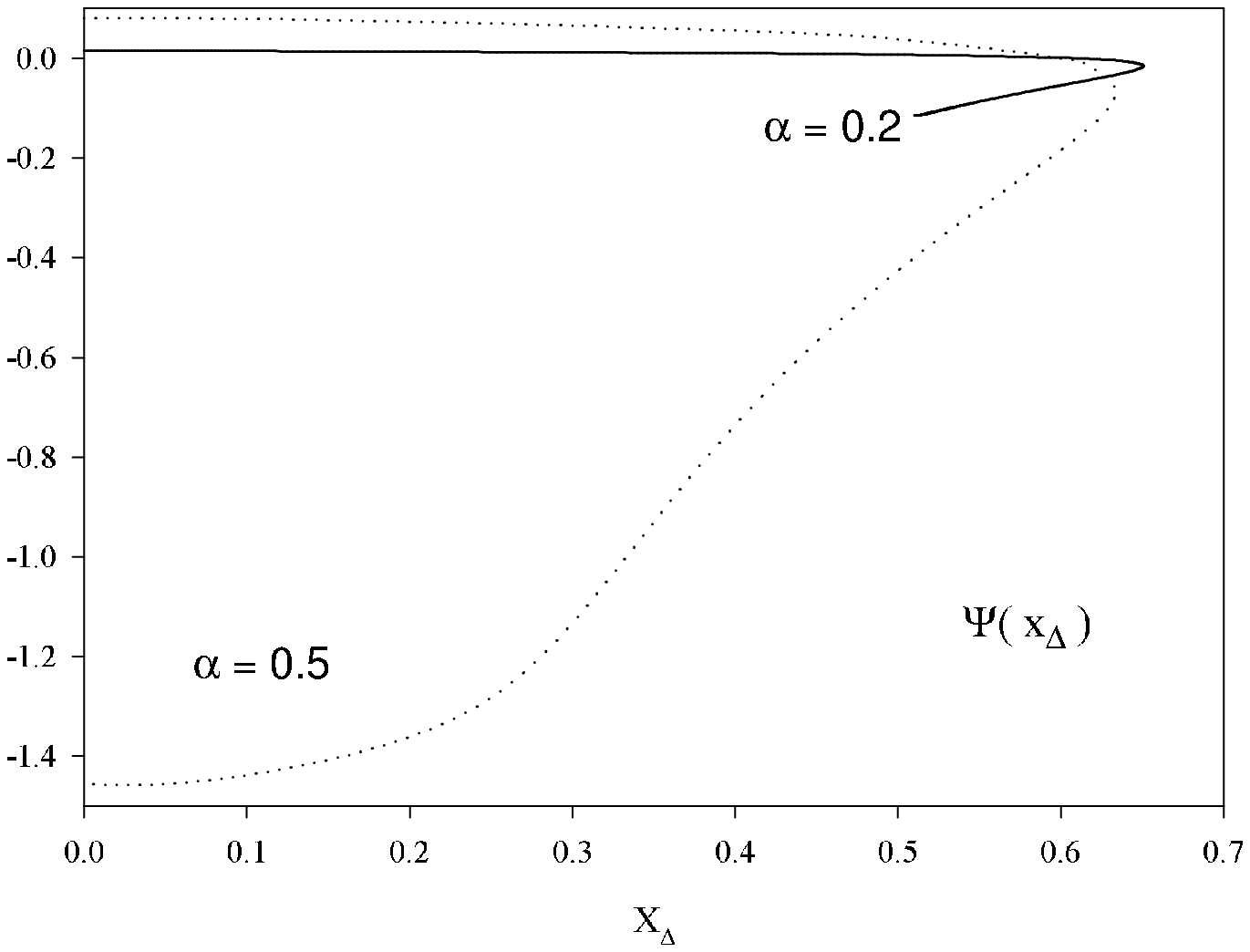,width=16cm}}
\end{picture}
\\
\\
\textbf{Figure 6.} The value of the metric function $ \psi$ of the uniform
black string ($m=0,~n=1$)  at the horizon $ \psi(x_{\Delta})$, is
shown as function of $x_{\Delta}$ for $\alpha=0.2$ and $\alpha=0.5$.
The upper and lower curves correspond to the 1., respectively 2. branch of solutions. 
\\
\\
\newpage
\setlength{\unitlength}{1cm}

\begin{picture}(16,16)
\centering
\put(0,0){\epsfig{file=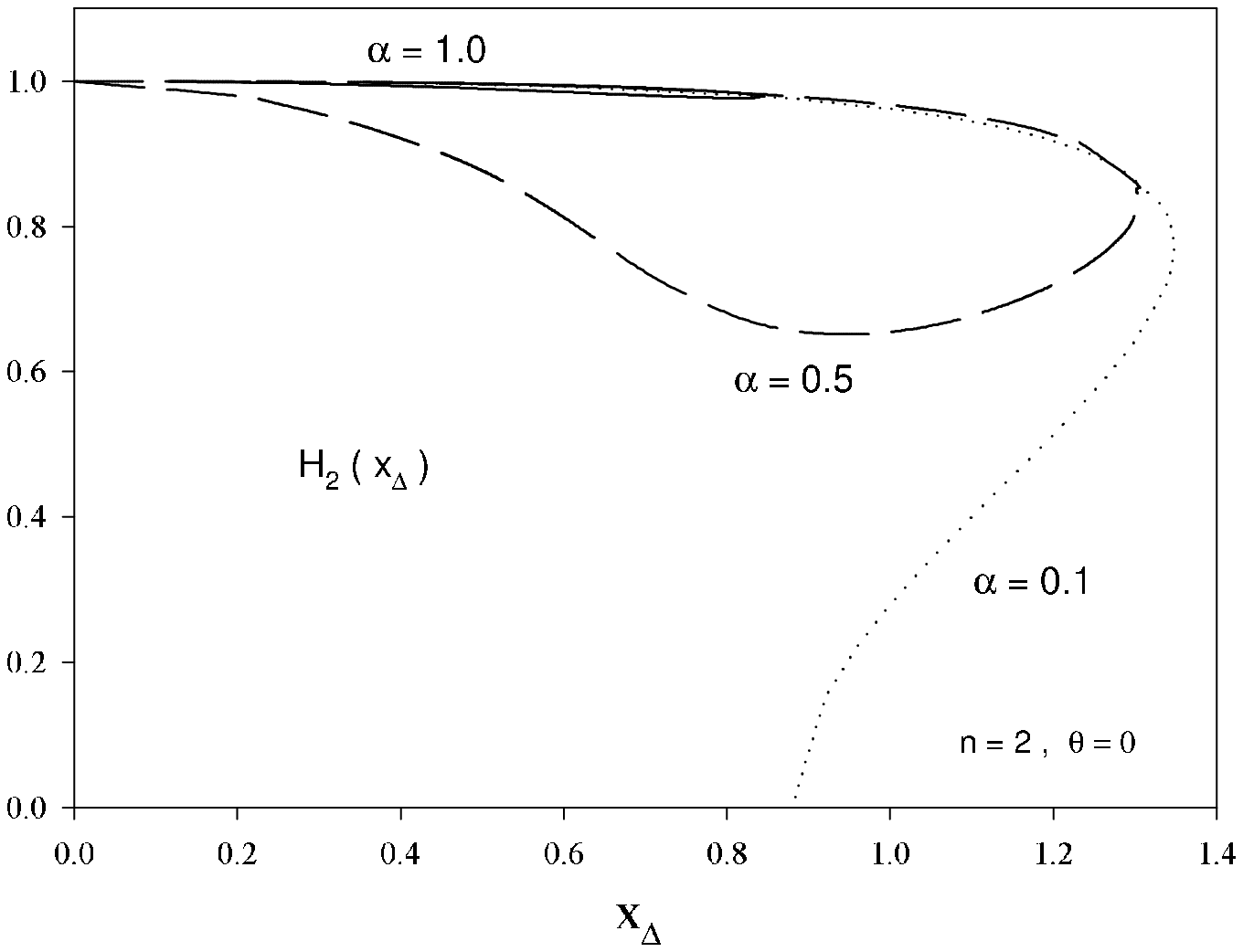,width=16cm}}
\end{picture}
\\
\\
\textbf{Figure 7.} 
 The value of the gauge field function $H_2$ of the
deformed black string ($m=0,~n=2$) at the horizon, $H_2(x_{\Delta},\theta=0)$
is shown as function of $x_{\Delta}$ for $\alpha=0.1$ and $\alpha=0.5$.
The upper and lower curves correspond to the 1., respectively 2. branch of solutions. 
Here and in Figures 8, 9 the curves for different
$\theta$ are essentially equal to those for $\theta=0$.

\newpage
\setlength{\unitlength}{1cm}

\begin{picture}(16,16)
\centering
\put(0,0){\epsfig{file=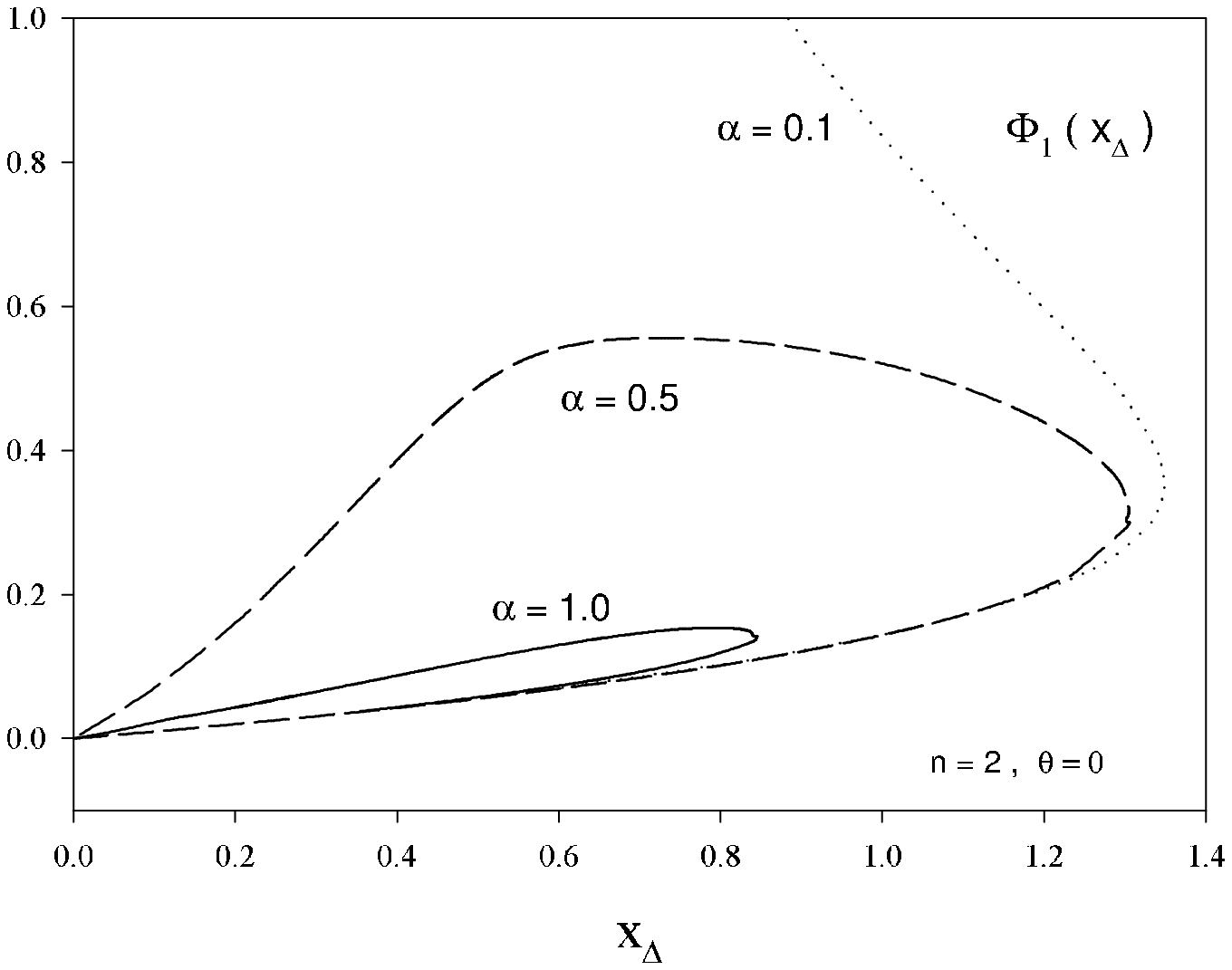,width=16cm}}
\end{picture}
\\
\\
\textbf{Figure 8.}
The value of the gauge field function $\Phi_1$ of the
deformed black string ($m=0,~n=2$) at the horizon, $\Phi_1(x_{\Delta},\theta=0)$
is shown as function of $x_{\Delta}$ for 
$\alpha=0.1$, $\alpha=0.5$ and $\alpha=1.0$.
The lower and upper curves correspond to the 1., respectively 2. branch of solutions. 
\newpage
\setlength{\unitlength}{1cm}

\begin{picture}(16,16)
\centering
\put(0,0){\epsfig{file=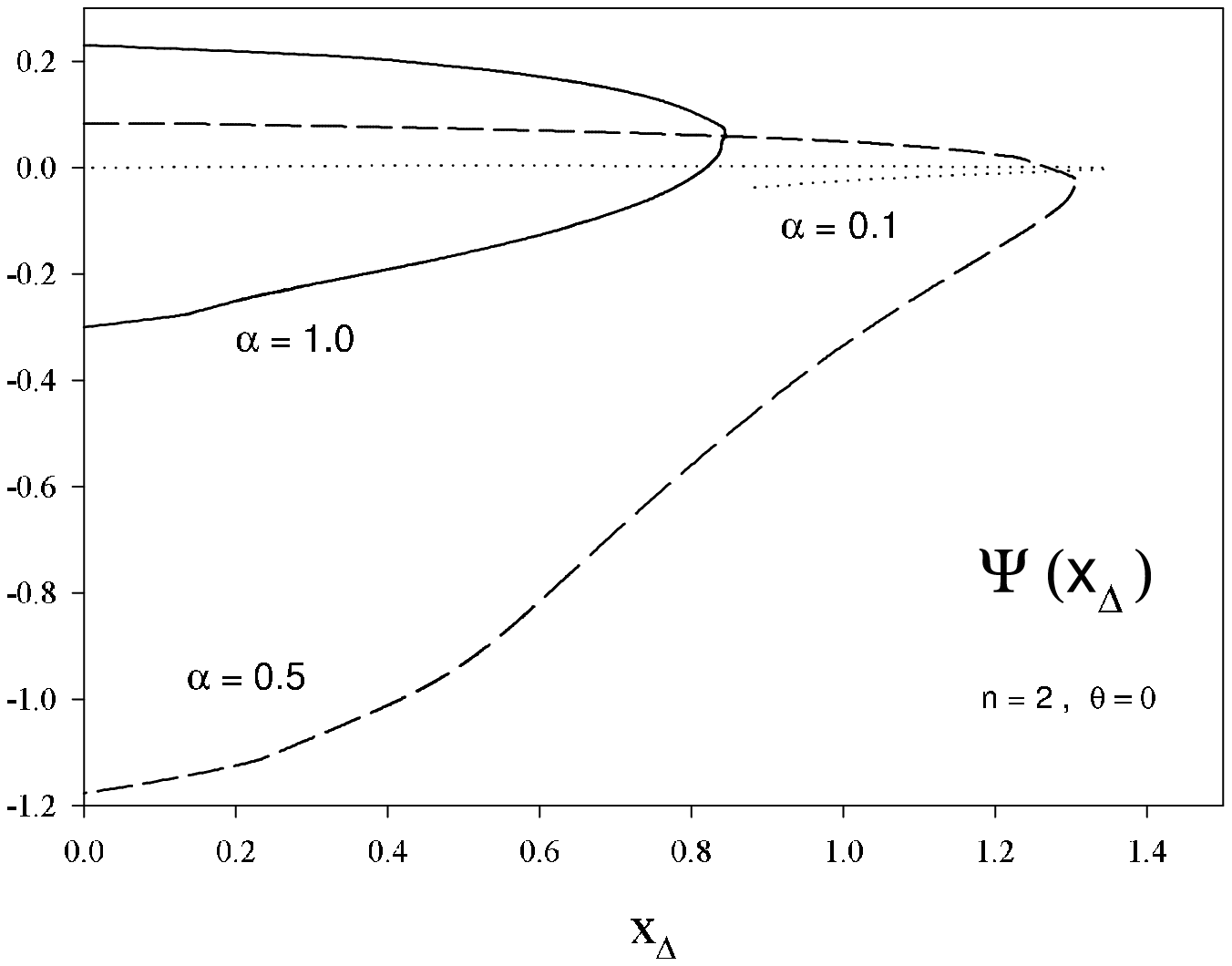,width=16cm}}
\end{picture}
\\
\\ 
\textbf{Figure 9.} The value of the metric function $ \psi$ of the
deformed black string ($m=0,~n=2$) at the horizon $\psi(x_{\Delta},\theta=0)$
is shown as function of $x_{\Delta}$ for $\alpha=0.1$, 
$\alpha=0.5$ and $\alpha=1.0$.
The upper and lower curves correspond to the 1., respectively 2. branch of solutions. 
\\
\\
\newpage
\setlength{\unitlength}{1cm}

\begin{picture}(16,16)
\centering
\put(0,0){\epsfig{file=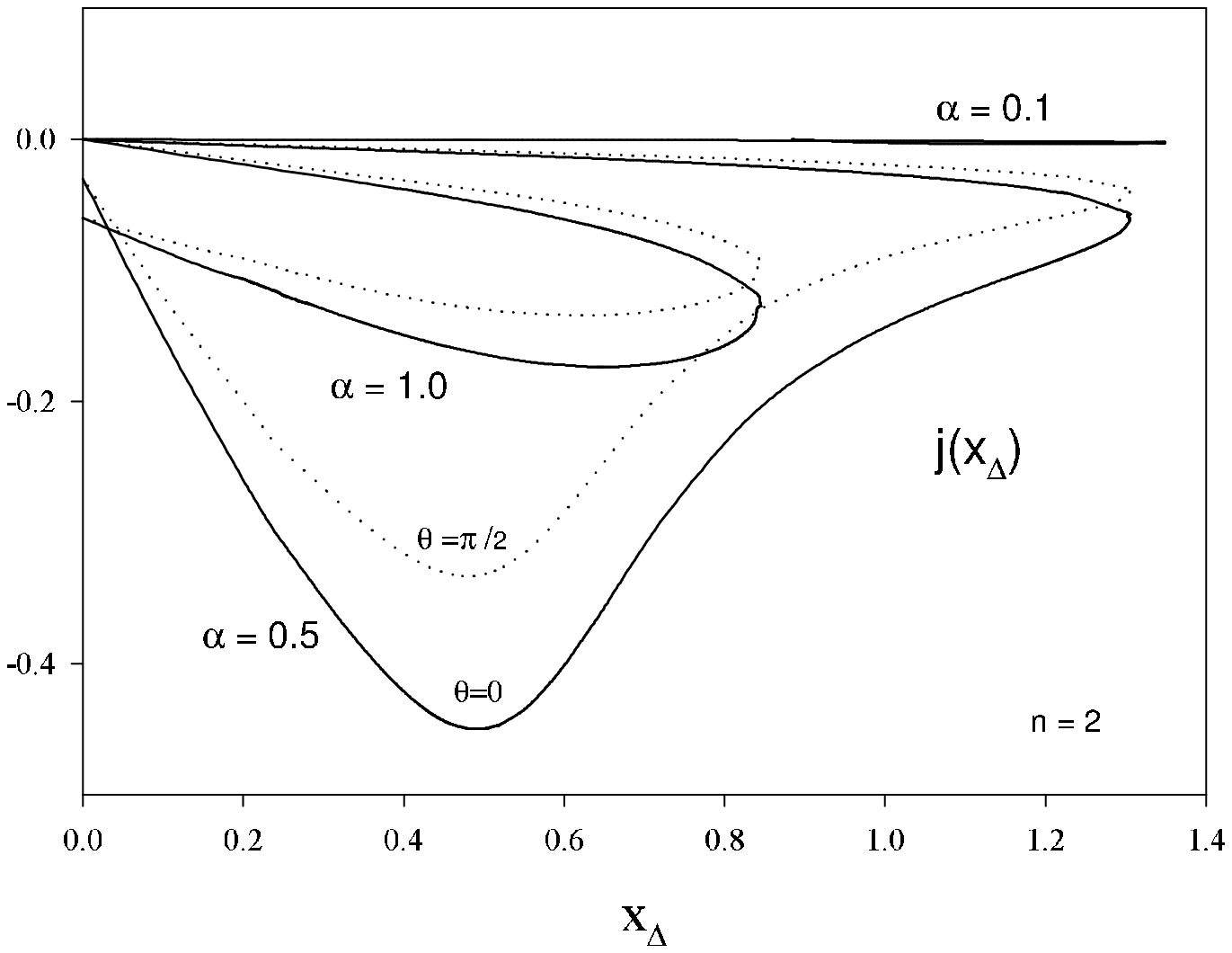,width=16cm}}
\end{picture}
\\
\\
\textbf{Figure 10. }
The value of the metric function $j$ used in the Ansatz described in  Appendix B of the
deformed black string ($m=0,~n=2$) at the horizon, $j(x_{\Delta})$
is shown as function of $x_{\Delta}$ for $\alpha=0.1$, $\alpha=0.5$ 
and $\alpha=1.0$.
We show the curves for $\theta=0$ (solid) and $\theta=\pi/2$ (dotted).
The upper and lower curves correspond to the 1., respectively 2. branch of solutions. 
\newpage
\setlength{\unitlength}{1cm}
\begin{picture}(16,16)
\centering
\put(0,0){\epsfig{file=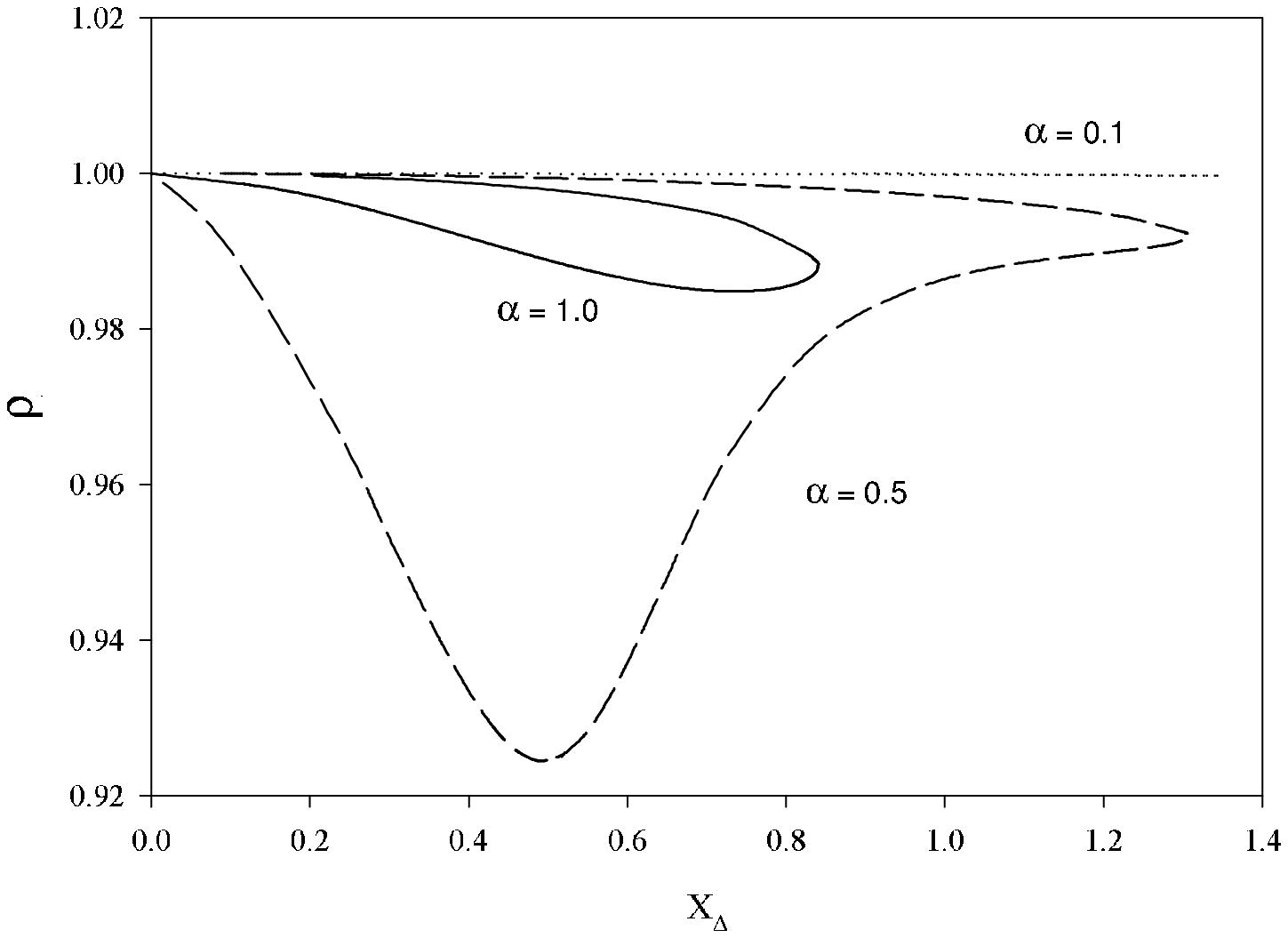,width=16cm}}
\end{picture}
\\
\\
\textbf{Figure 11.} 
The ratio $\rho=L_e/L_p$ of the horizon circumference along the equator
$L_e$ and along the poles $L_p$ is shown for the deformed
black strings ($m=0,~n=2$) as function
of the horizon parameter $x_{\Delta}$ 
for $\alpha=0.1$, $\alpha=0.5$ and $\alpha=1.0$.
The upper and lower curves correspond to the 1., respectively 2. branch of solutions. 
\newpage
\setlength{\unitlength}{1cm}

\begin{picture}(16,16)
\centering
\put(0,0){\epsfig{file=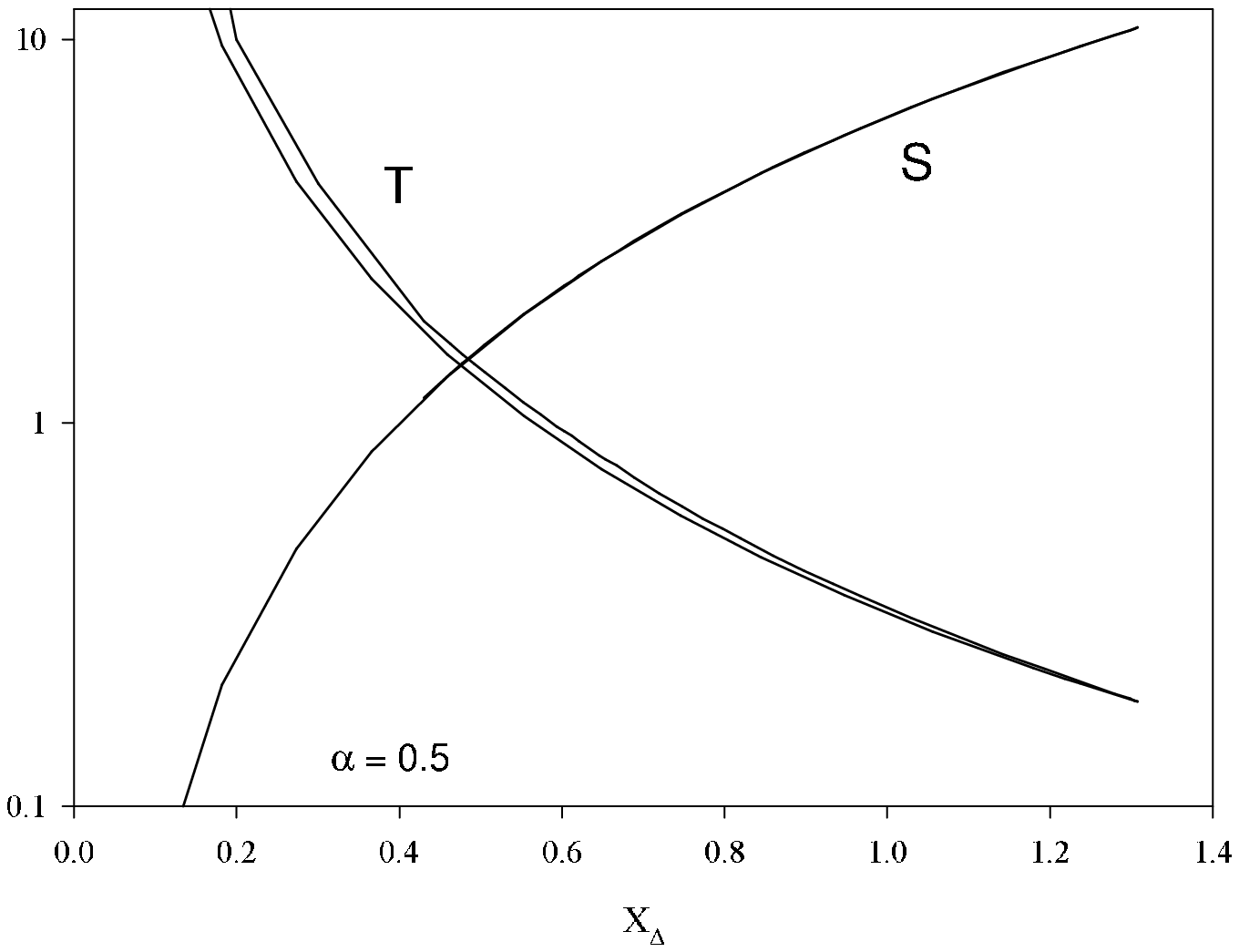,width=16cm}}
\end{picture}
\\
\\
\textbf{Figure 12.} The temperature $T$ and the entropy 
$S$ of the deformed black strings ($m=0,~n=2$)
are shown as functions of $x_{\Delta}$ for $\alpha=0.5$

\newpage

\setlength{\unitlength}{1cm}

\begin{picture}(16,16)
\centering
\put(0,0){\epsfig{file=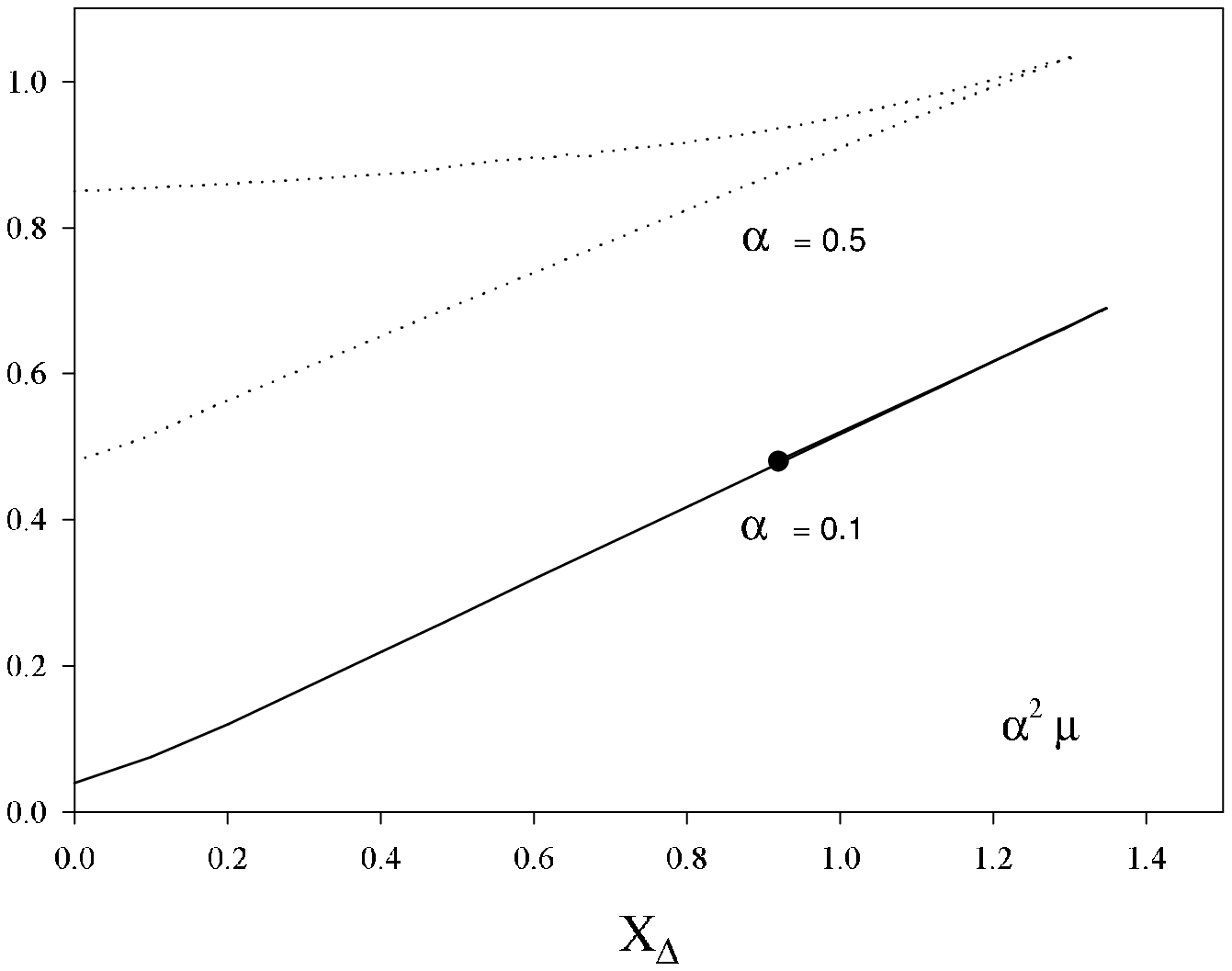,width=16cm}}
\end{picture}
\\
\\
\textbf{Figure 13.} The mass of the $(m=0,~n=2)$ deformed black string solutions 
is shown as function of $x_{\Delta}$ for  $\alpha=0.1$ and $\alpha=0.5$.
On the $\alpha=0.1$ plot, the bullet shows where the second branch stops.
 The lower and upper curves correspond to the 1., respectively 2. branch of solutions.   
\newpage

\setlength{\unitlength}{1cm}

\begin{picture}(16,16)
\centering
\put(0,0){\epsfig{file=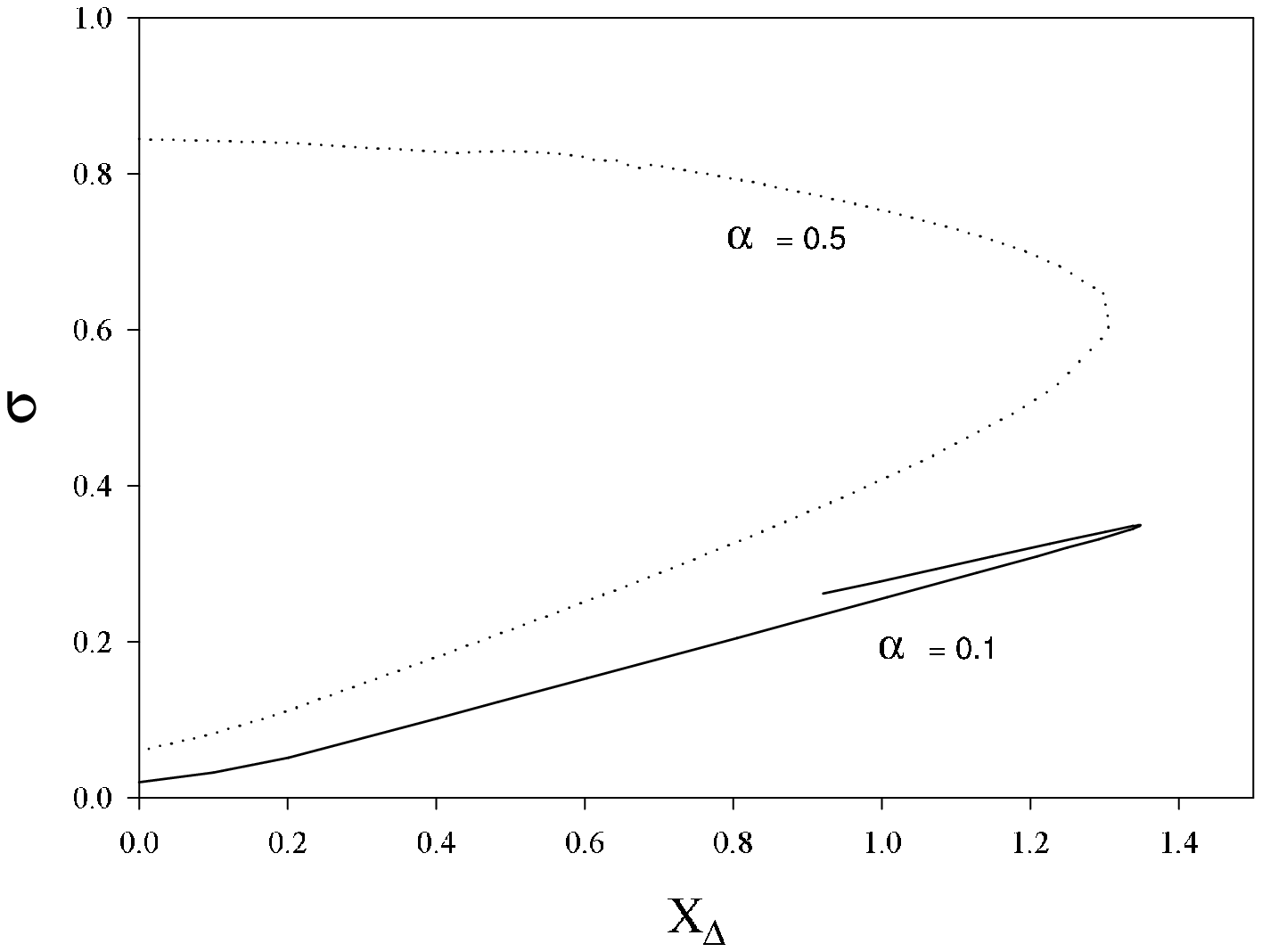,width=16cm}}
\end{picture}
\\
\\
\textbf{Figure 14.} The tension $\sigma$ of the $(m=0,~n=2)$
deformed  black string solutions 
is shown as function of $x_{\Delta}$ for  $\alpha=0.1$ and $\alpha=0.5$.
The lower and upper curves correspond to the 1., respectively 2. branch of solutions. 

\newpage

\setlength{\unitlength}{1cm}

\begin{picture}(16,16)
\centering
\put(0,0){\epsfig{file=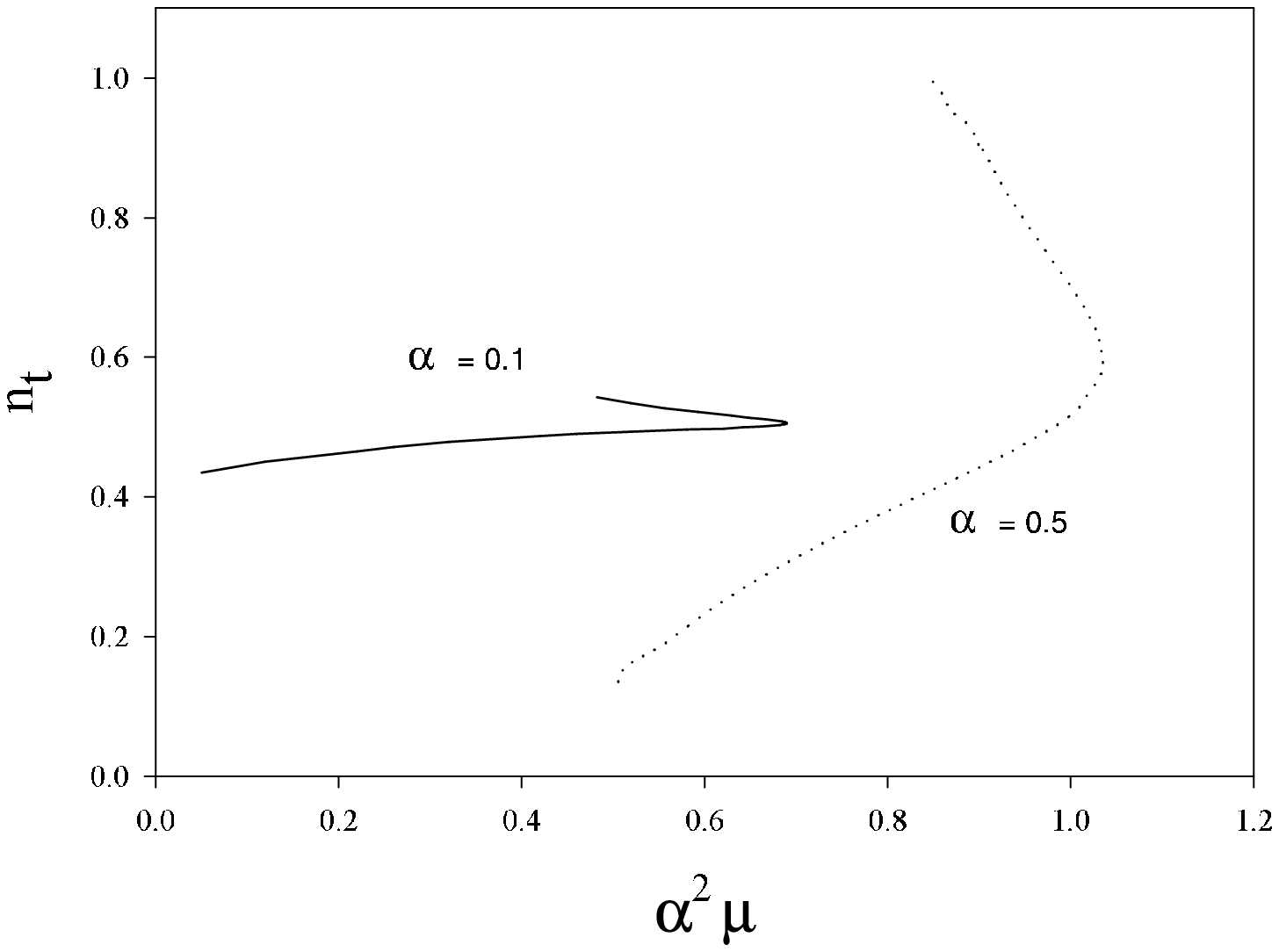,width=16cm}}
\end{picture}
\\
\\
\textbf{Figure 15.} A ($\mu,n_t$) diagram is plotted for $(m=0,~n=2)$
deformed  black string solutions 
and two values of $\alpha $.
\newpage
\setlength{\unitlength}{1cm}
 \begin{picture}(16,16)
 \centering
 \put(0,0){\epsfig{file=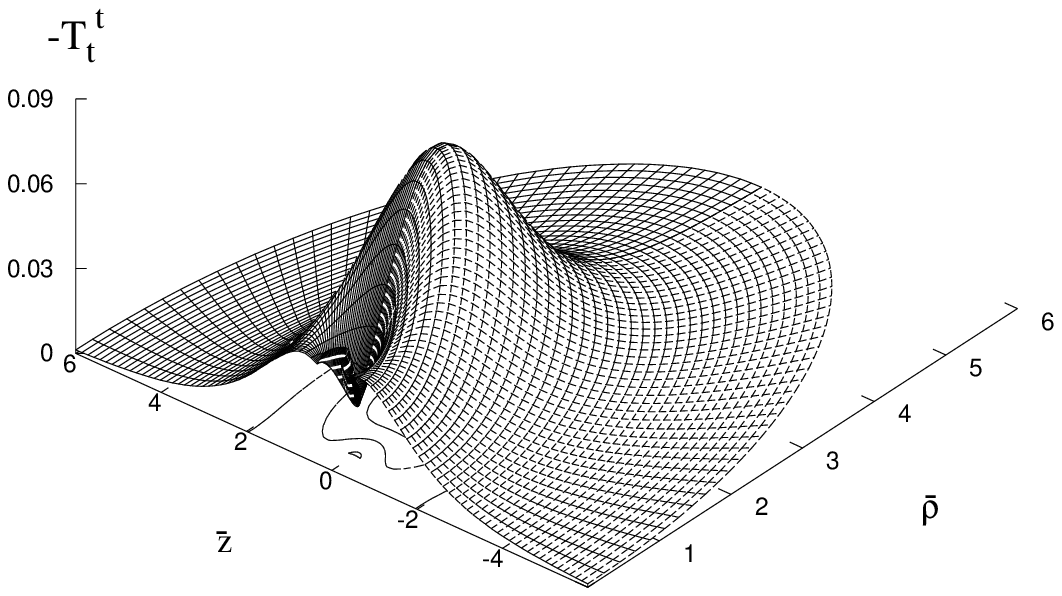,width=16cm}}
 \end{picture}
 \\
 \\
\textbf{Figure 16.} 
The energy density of the matter fields $\epsilon=-T_t^t$
is shown as a function
of the coordinates $\bar{\rho}=r\sin \theta$, $\bar{z}=r \cos \theta$ for a
$(m=0,~n=2)$
deformed black string solution with $\alpha=0.5,~r_h=0.2$.
 


%

\newpage

\setlength{\unitlength}{1cm}

\begin{picture}(16,16)
\centering
\put(0,0){\epsfig{file=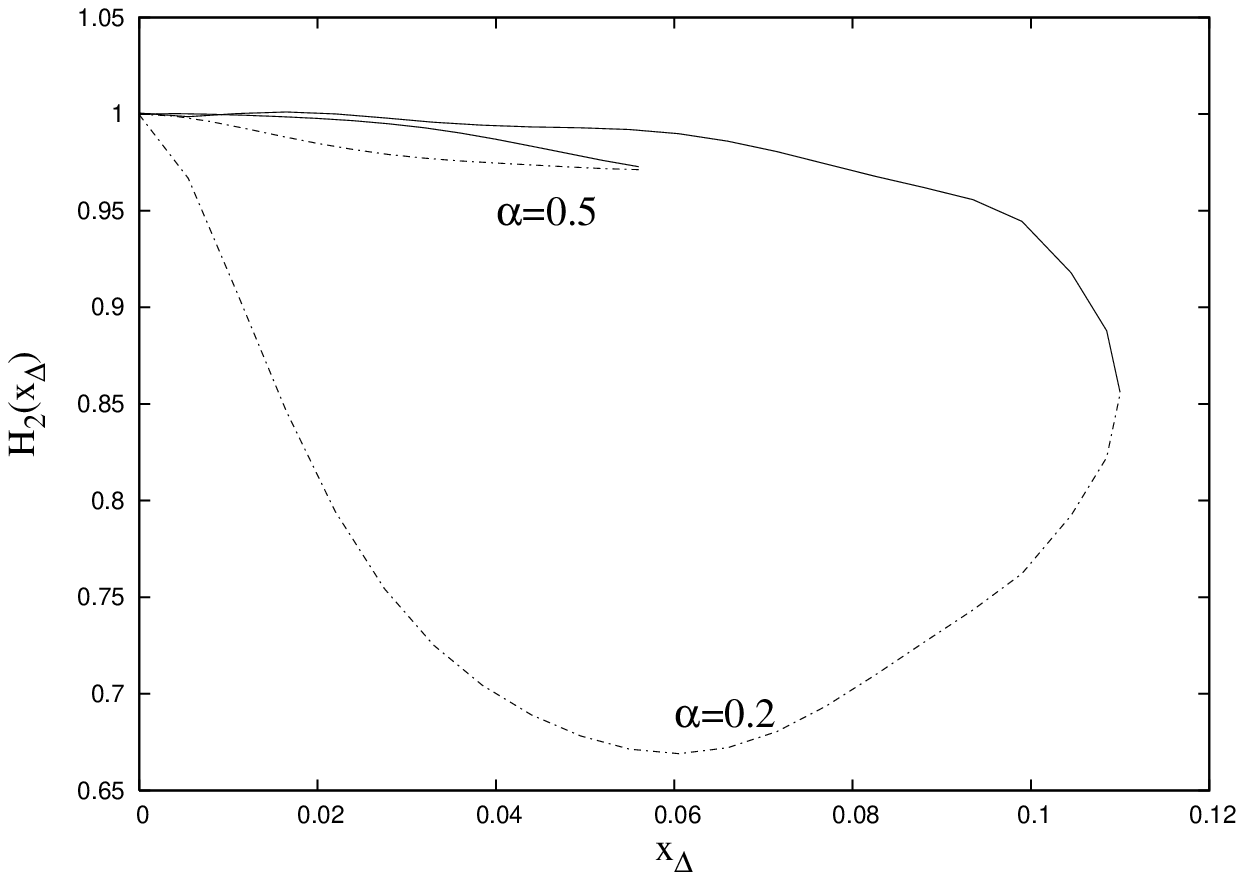,width=16cm}}
\end{picture}
\\
\\
\textbf{Figure 17.}  The value of the gauge field function $H_2$  
at the horizon, $H_2(x_{\Delta},\theta=0)$
is shown as function of $x_{\Delta}$
for $(m=1,~n=1)$ black string solutions with  $\alpha=0.2$ and $\alpha=0.5$.
Here and in Figures 18-25, the dotted line denotes the higher branch solution,
the continuous line corresponding to the fundamental branch.

\newpage
\setlength{\unitlength}{1cm}

\begin{picture}(16,16)
\centering
\put(0,0){\epsfig{file=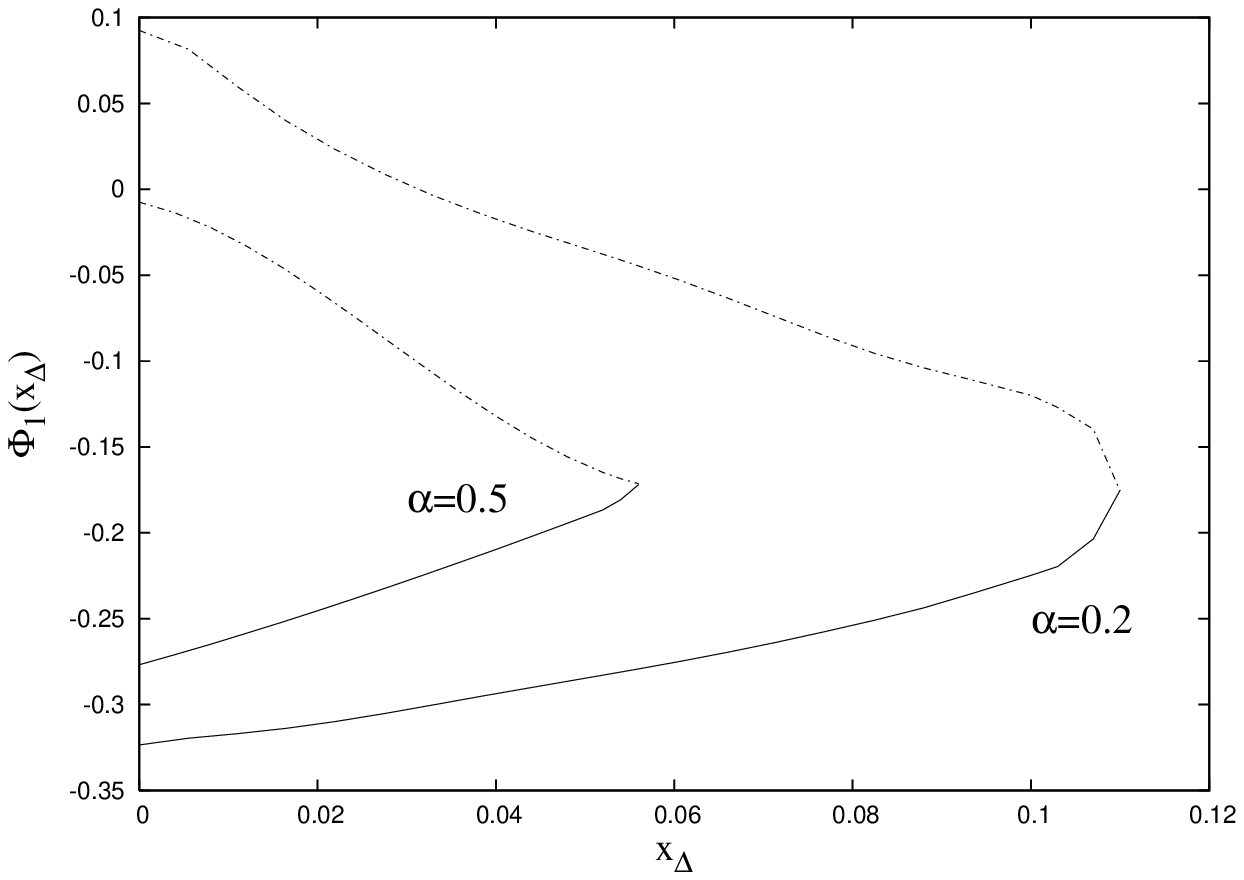,width=16cm}}
\end{picture}
\\
\\
\textbf{Figure 18.} 
The value of the gauge field function $\Phi_1$
at the horizon, $\Phi_2(x_{\Delta},\theta=0)$
is shown as function of $x_{\Delta}$
for $(m=1,~n=1)$ black string solutions  with $\alpha=0.2$ and $\alpha=0.5$.
 
\newpage
\setlength{\unitlength}{1cm}

\begin{picture}(16,16)
\centering
\put(0,0){\epsfig{file=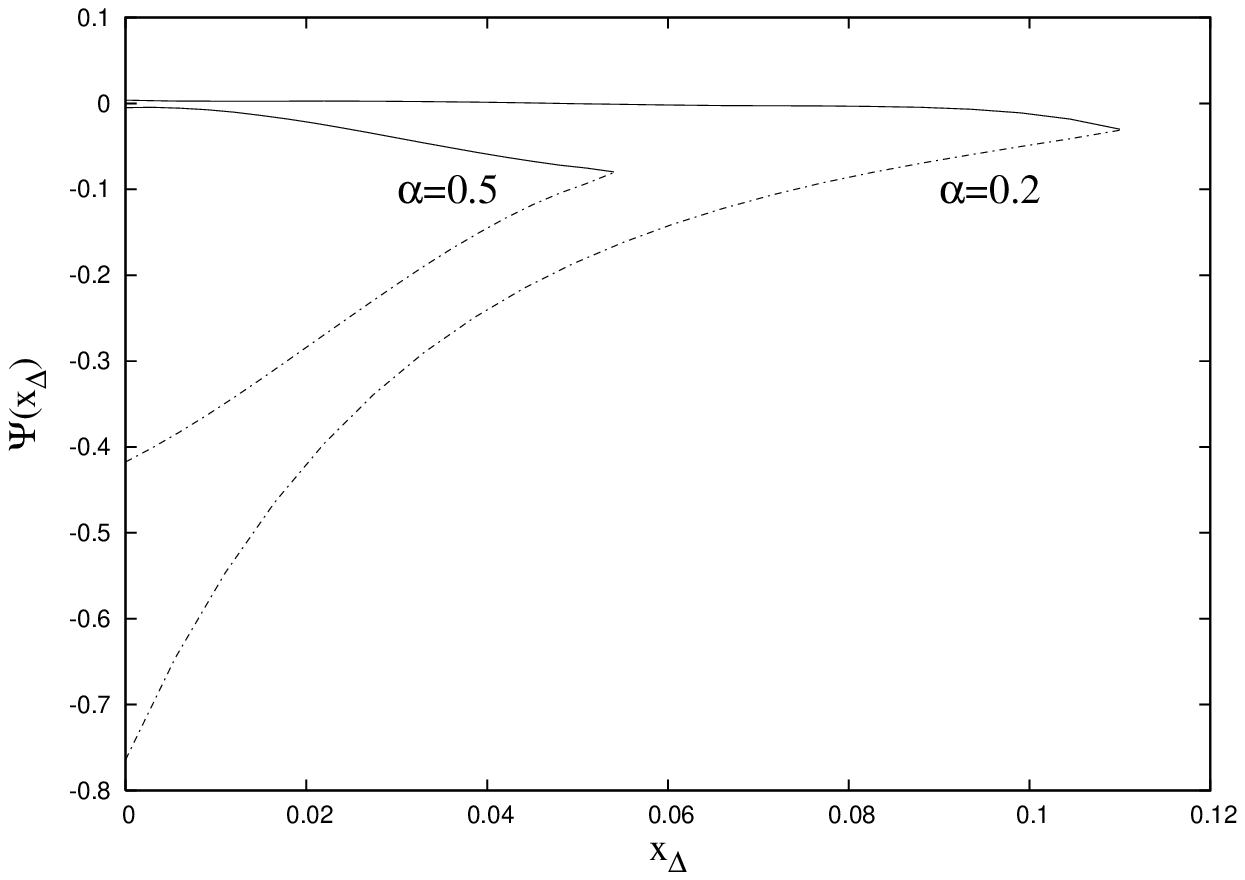,width=16cm}}
\end{picture}
\\
\\
\textbf{Figure 19.} The value of the metric function $\psi$ 
 at the horizon $\psi(x_{\Delta},\theta=0)$
is shown as function of $x_{\Delta}$ 
for $(m=1,~n=1)$ black string solutions with $\alpha=0.2$ and $\alpha=0.5$.  

\newpage

\setlength{\unitlength}{1cm}

\begin{picture}(16,16)
\centering
\put(0,0){\epsfig{file=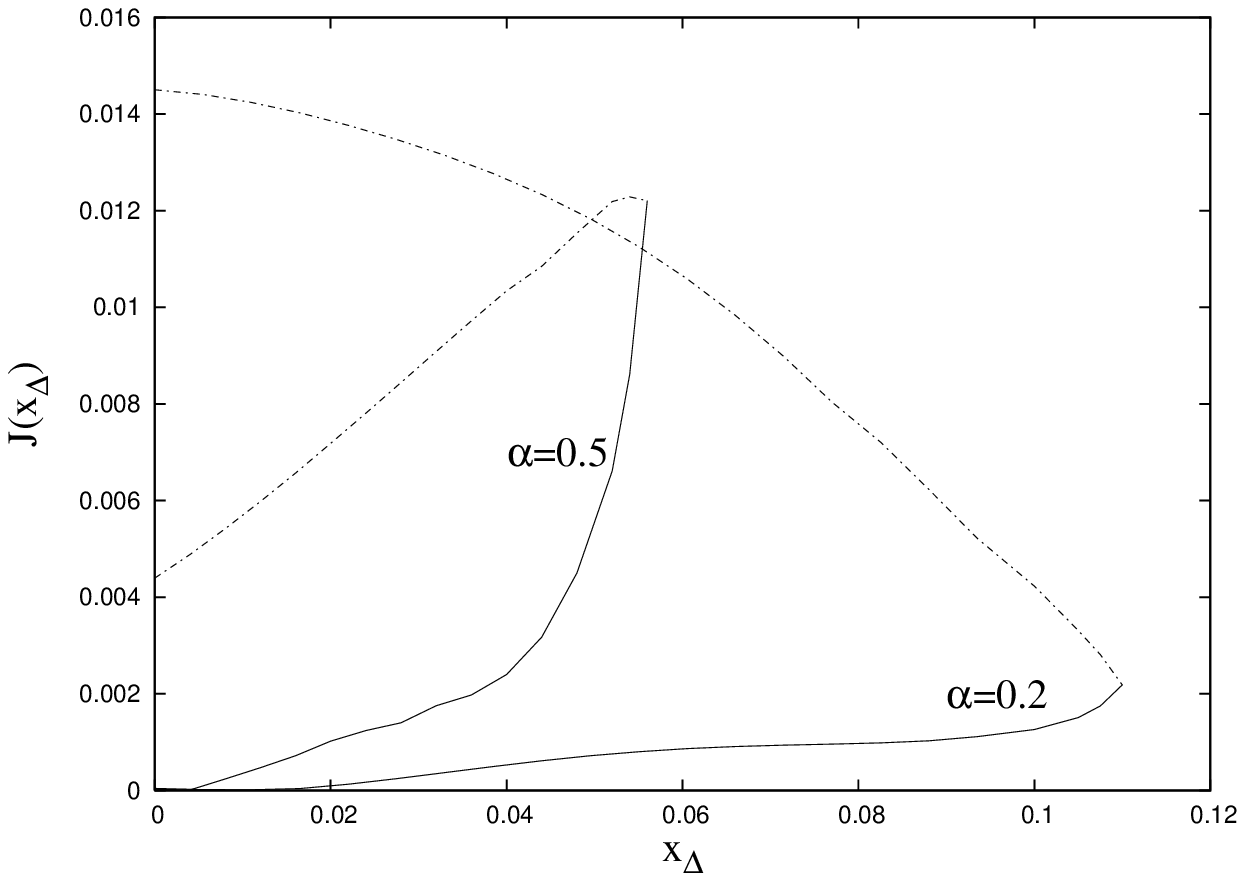,width=16cm}}
\end{picture}
\\
\\
\textbf{Figure 20.} The value of the metric  function $J$  
 at the horizon, $J(x_{\Delta},\theta=\pi/2)$
is shown as function of $x_{\Delta}$ 
for $(m=1,~n=1)$ black string solutions with $\alpha=0.2$ and $\alpha=0.5$.  

\newpage

\setlength{\unitlength}{1cm}

\begin{picture}(16,16)
\centering
\put(0,0){\epsfig{file=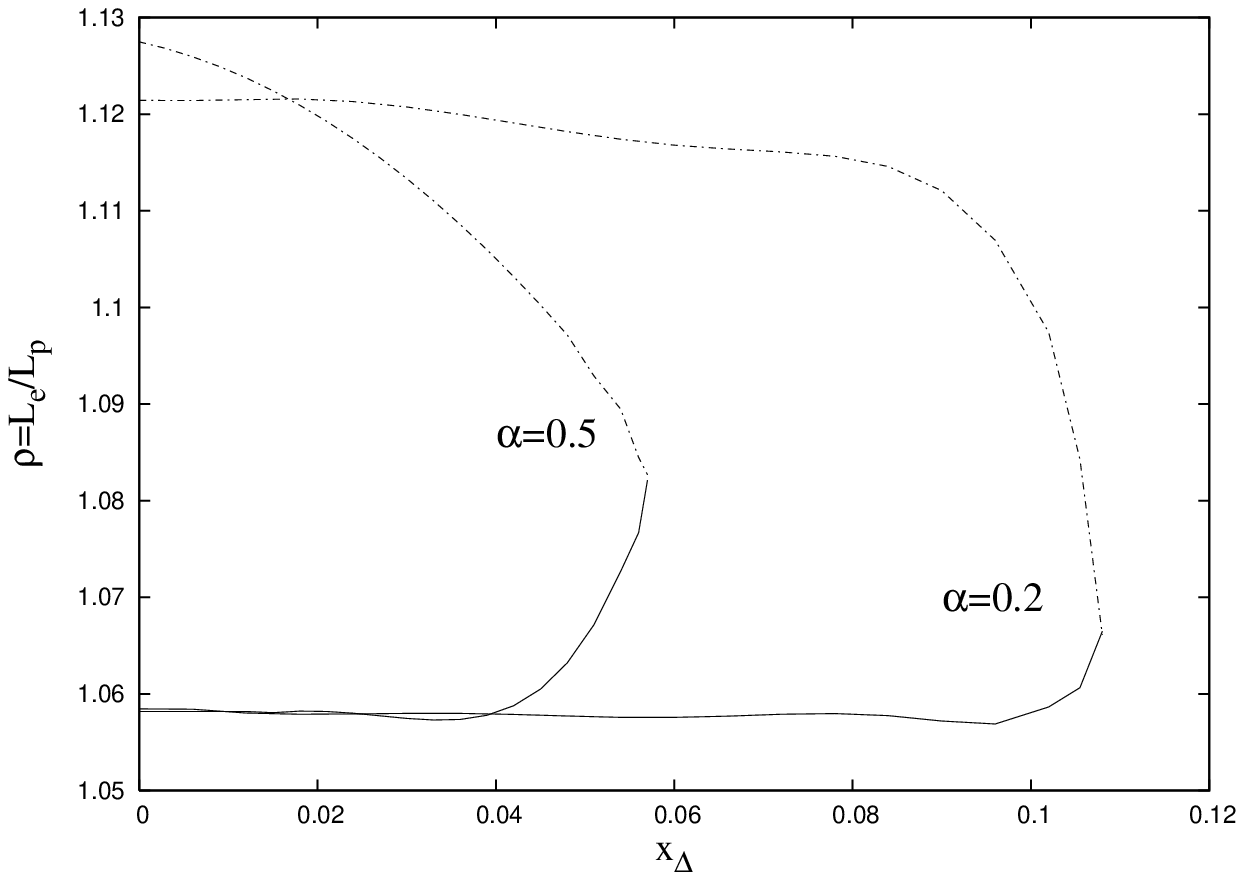,width=16cm}}
\end{picture}
\\
\\
\textbf{Figure 21.} The ratio $\rho=L_e/L_p$ of the horizon circumference along the equator
$L_e$ and along the poles $L_p$ is shown for $(m=1,~n=1)$ black string 
solutions  as function
of the horizon parameter $x_{\Delta}$ 
for $\alpha=0.2$ and $\alpha=0.5$.
 
\newpage

\setlength{\unitlength}{1cm}

\begin{picture}(6,10)
\centering
\put(0,0){\epsfig{file=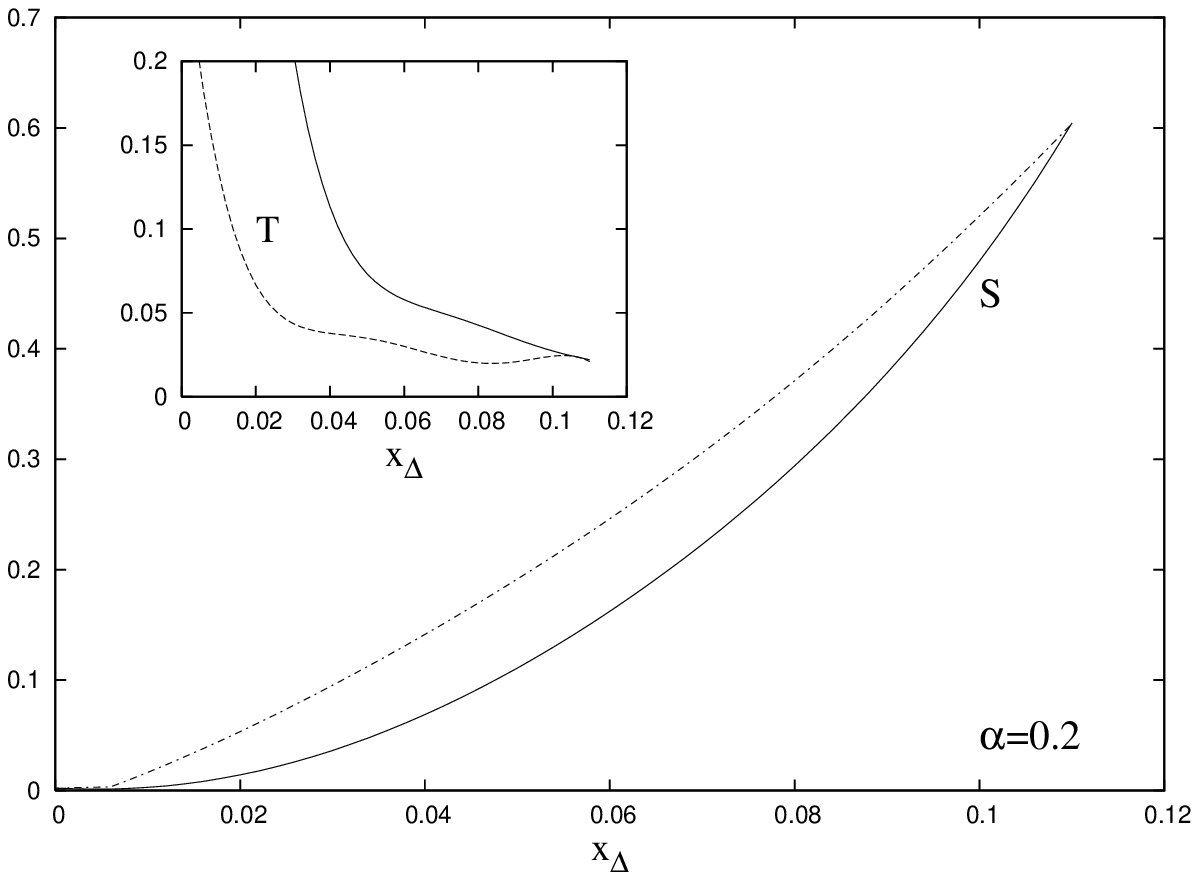,width=14cm}}
\end{picture}
 
\begin{picture}(22,10)
\centering
\put(0.,0){\epsfig{file=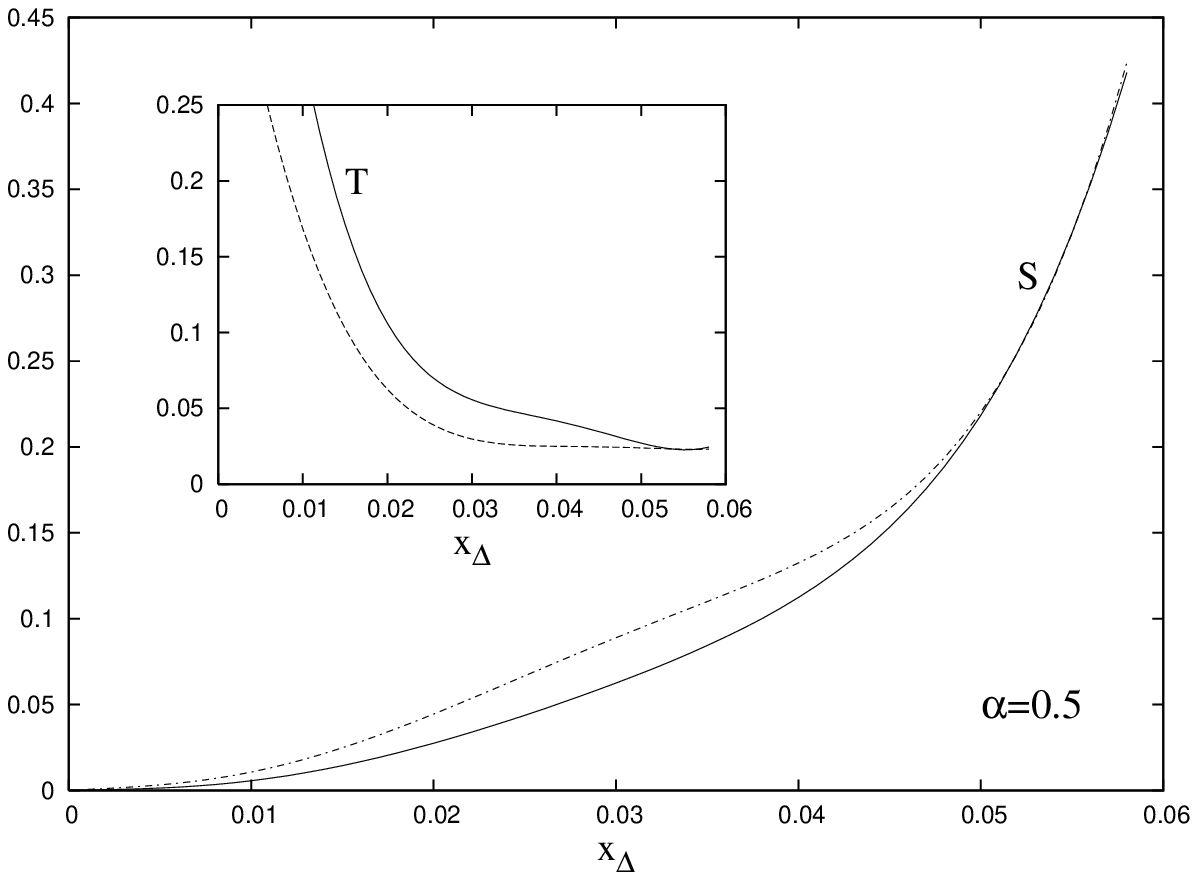,width=14cm}}
\end{picture}
\\
\\
\textbf{Figure 22.} 
The temperature $T$ and the  entropy $S$  
are showns as functions of $x_{\Delta}$ for $(m=1,~n=1)$ black string 
solutions with $\alpha=0.2$  and $\alpha=0.5$.
\newpage

\setlength{\unitlength}{1cm}

\begin{picture}(16,16)
\centering
\put(0,0){\epsfig{file=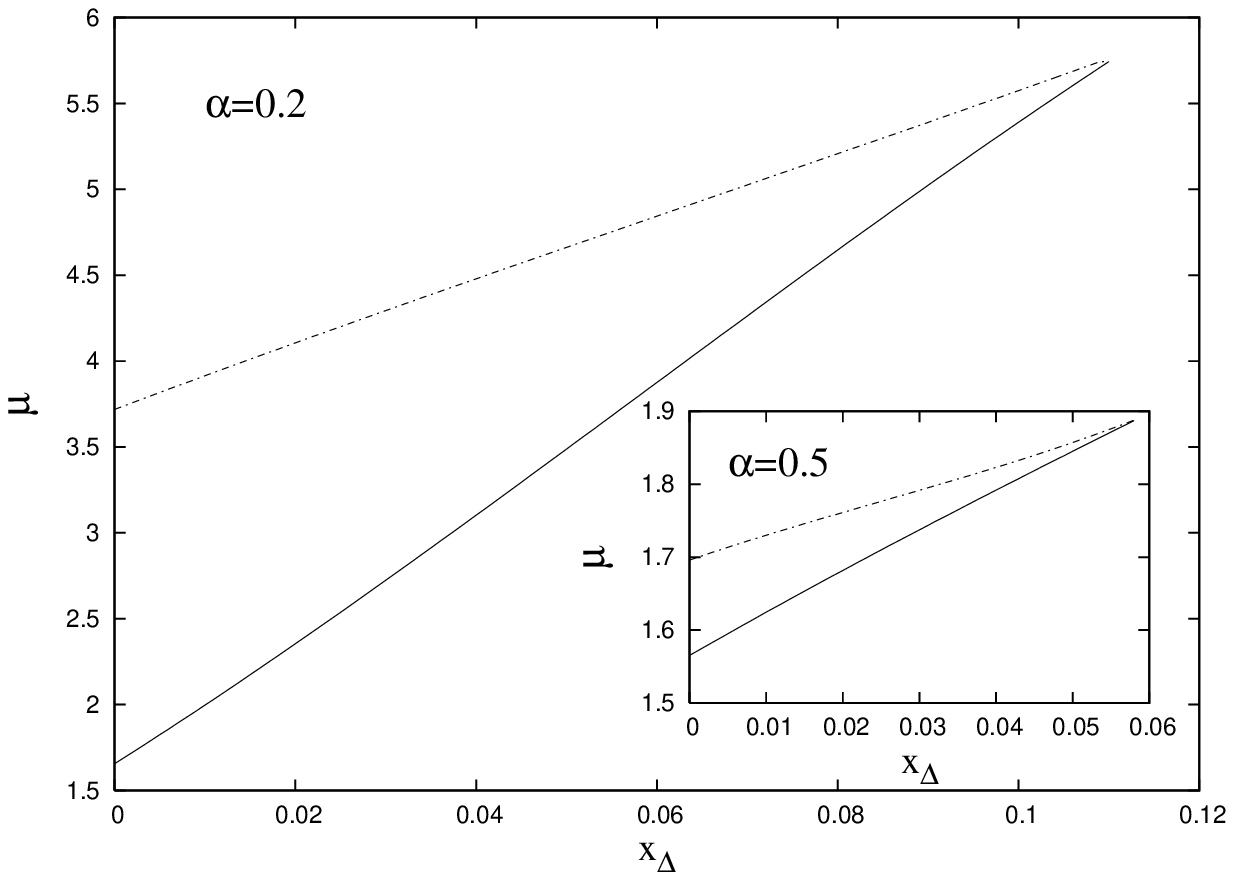,width=16cm}}
\end{picture}
\\
\\
\textbf{Figure 23.} The mass of the $(m=1,~n=1)$ black string solutions 
is shown as function of $x_{\Delta}$ for  $\alpha=0.2$ and $\alpha=0.5$.

\newpage

\setlength{\unitlength}{1cm}

\begin{picture}(16,16)
\centering
\put(0,0){\epsfig{file=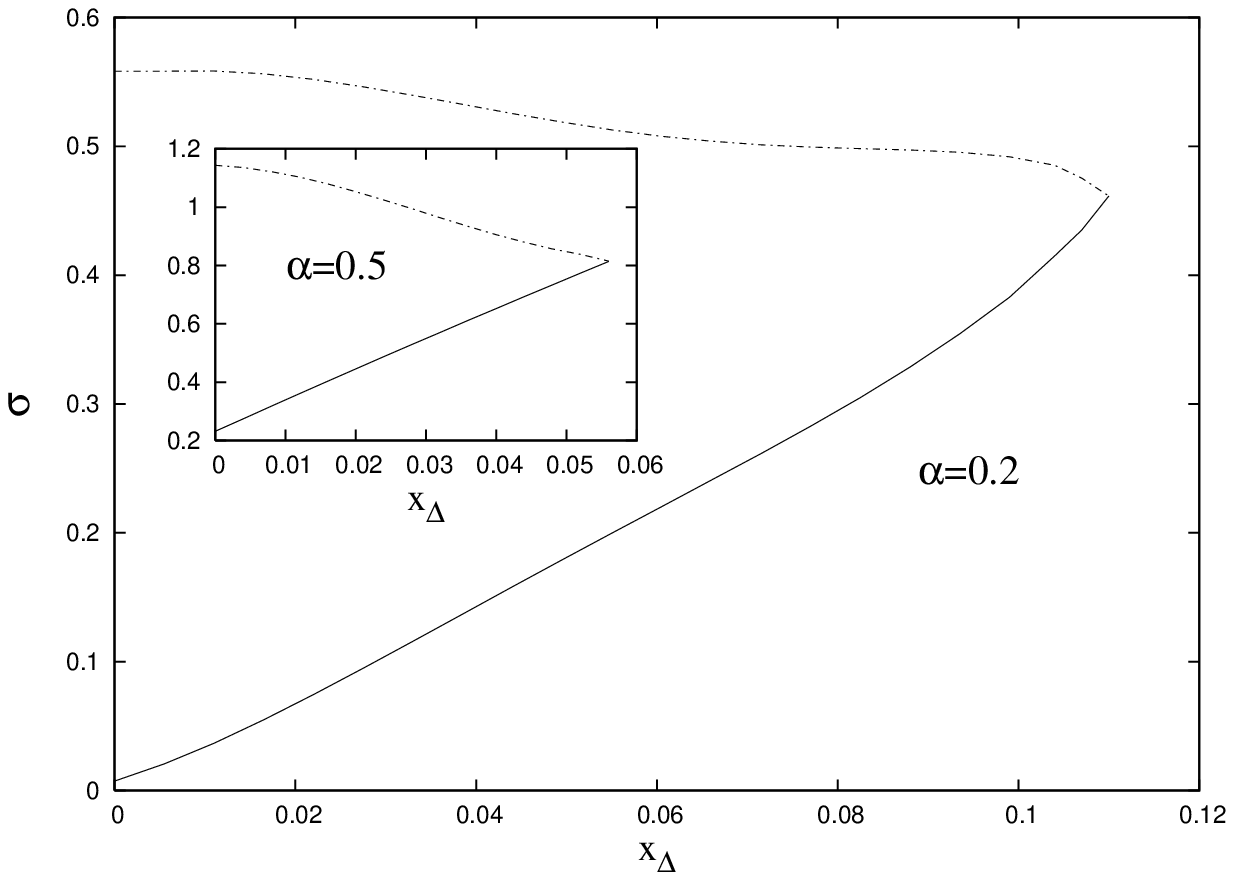,width=16cm}}
\end{picture}
\\
\\
\textbf{Figure 24.} The tension $\sigma$ of the $(m=1,~n=1)$ black string solutions 
is shown as function of $x_{\Delta}$ for  $\alpha=0.2$ and $\alpha=0.5$.
}

\newpage

\setlength{\unitlength}{1cm}

\begin{picture}(16,16)
\centering
\put(0,0){\epsfig{file=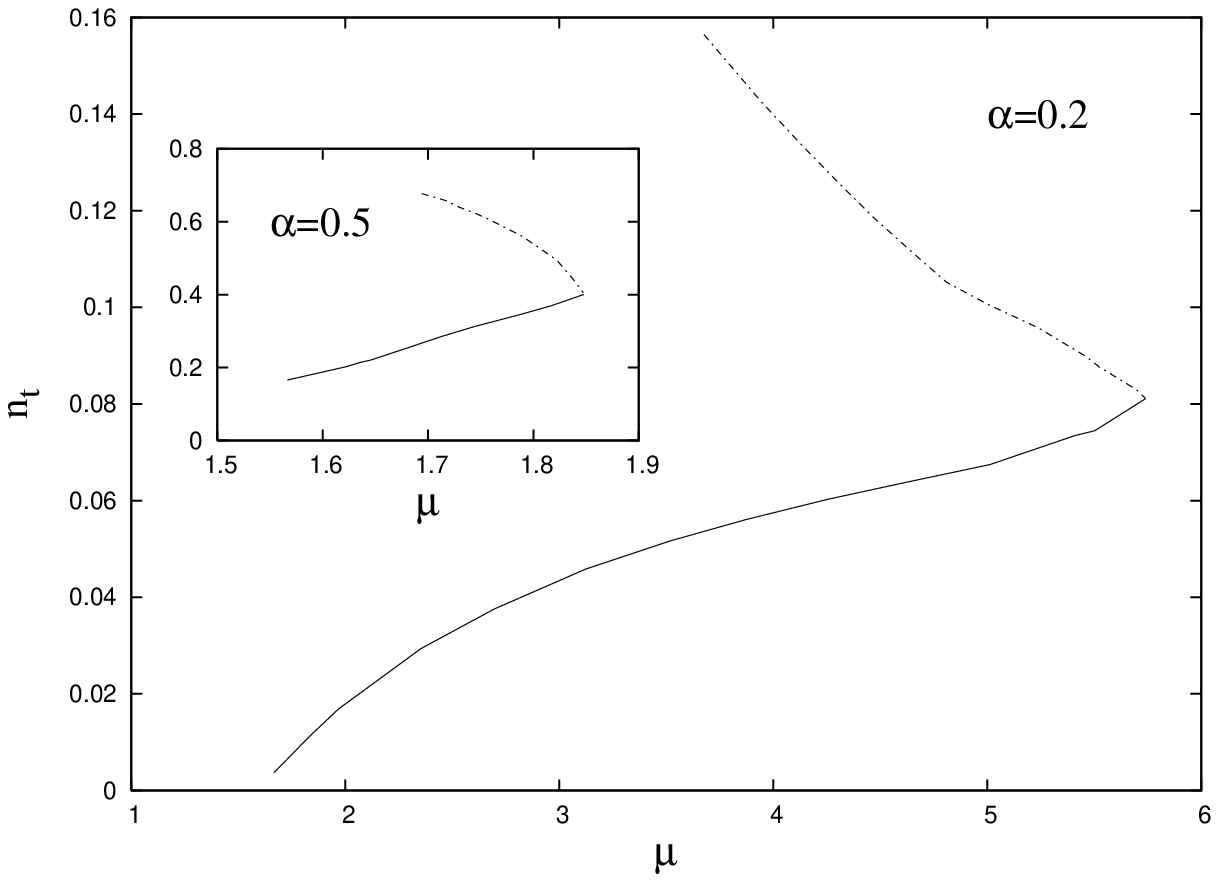,width=16cm}}
\end{picture}
\\
\\
\textbf{Figure 25.} A ($\mu,n_t$) diagram is plotted for $(m=1,~n=1)$ black string solutions 
and two values of $\alpha $. 


\newpage
\setlength{\unitlength}{1cm}
\begin{picture}(16,16)
\centering
\put(0,0){\epsfig{file=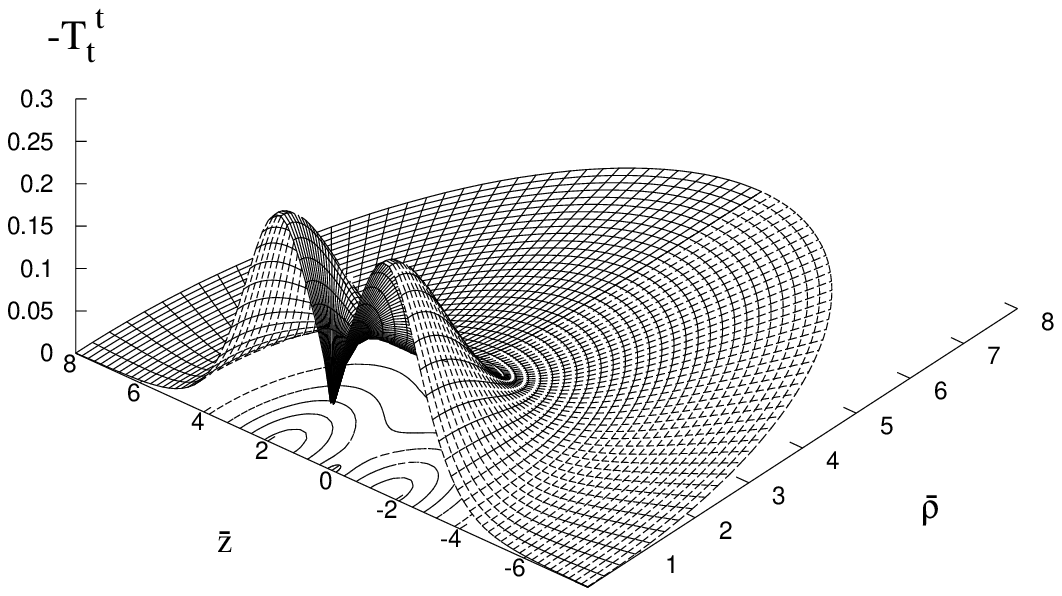,width=16cm}}
\end{picture}
 \\
 \\
\textbf{Figure 26.} The energy density of the matter fields $\epsilon=-T_t^t$
is shown as a function
of the coordinates $\bar{\rho}=r\sin \theta$, $\bar{z}=r \cos \theta$ for a $(m=1,~n=1)$
lower branch black string solution with $\alpha=0.2$ $r_h=0.04$.
 
\end{document}